\documentclass[10pt,journal,final]{IEEEtran}

\usepackage{graphicx}
\graphicspath{{./Figures/}}
\usepackage{subfigure} 
\usepackage{xcolor,graphicx,float} 

\usepackage{amsmath}
\usepackage{multirow}
\usepackage{amssymb}
\usepackage{algorithmic}
\usepackage{algorithm}
\usepackage{booktabs}
\usepackage{cite}
\usepackage[colorlinks,linkcolor=blue,citecolor=blue,urlcolor=black]{hyperref}
\usepackage{url}

\usepackage{titlesec}

\begin{document}
\title{Coordinating Flexible Demand Response and Renewable Uncertainties for Scheduling of Community Integrated Energy Systems with an Electric Vehicle Charging Station: A Bi-level Approach}

\author{Yang~Li,~\IEEEmembership{Senior Member,~IEEE,} Meng~Han, Zhen~Yang, Guoqing~Li 
\thanks{Y. Li, M. Han and G. Li are with the School of Electrical Engineering, Northeast Electric Power University, Jilin 132012, China (e-mail: liyang@neepu.edu.cn; 2322451905@qq.com; lgq@neepu.edu.cn).}%

\thanks{Z. Yang is with the State Grid Beijing Electric Power Company, Beijing 100032, China (e-mail: 1678084931@qq.com). }}

\markboth{IEEE TRANSACTIONS ON SUSTAINABLE ENERGY}{Y.~Li \MakeLowercase{\textit{et al.}}}

\maketitle
\begin{abstract}
   A community integrated energy system (CIES) with an electric vehicle charging station (EVCS) provides a new way for tackling growing concerns of energy efficiency and environmental pollution, it is a critical task to coordinate flexible demand response and multiple renewable uncertainties. To this end, a novel bi-level optimal dispatching model for the CIES with an EVCS in multi-stakeholder scenarios is established in this paper. In this model, an integrated demand response program is designed to promote a balance between energy supply and demand while maintaining a user comprehensive satisfaction within an acceptable range. To further tap the potential of demand response 
  through flexibly guiding users’ energy consumption and electric vehicles’ behaviors (charging,  discharging  and  providing spinning reserves), a dynamic pricing mechanism combining time-of-use and real-time pricing is put forward.
   In the solution phase, by using sequence operation theory (SOT), the original chance-constrained programming (CCP) model is converted into a readily solvable mixed-integer linear programming (MILP) formulation and finally solved by CPLEX solver. The simulation results on a practical CIES located in North China demonstrate that the presented method manages to balance the interests between CIES and EVCS via the coordination of flexible demand response and uncertain renewables.
\end{abstract}
    

\section{Introduction}
\IEEEPARstart{T}{oday}, the sustainable development of modern society has been threatened by the energy crisis and global warming. The efficient utilization of renewable generations (RGs) and the rapid development of electric vehicles (EVs) provide an opportunity to solve these two problems \cite{lv2019intelligent}. As an effective carrier of renewable energy resources, an integrated energy system (IES) can effectively integrate various types of distributed energy, loads, energy storage and other devices and control systems to satisfy multi-energy demand on the user side \cite{8494791}. The community integrated energy system (CIES) is a typical demonstration of the energy internet and will become a new model for the key development of a community in the future \cite{jin2021optimal}. Unfortunately, the inherent volatility and intermittency of RGs cause  large amounts of renewable curtailments \cite{7972993} and increase the difficulty of CIES scheduling \cite{wang2018coordinated}. Demand responses of flexible loads such as EVs have proven to be beneficial to address the problem of renewable uncertainty. However, with increasing penetration of EVs, the disorderly charging behavior of EVs will inevitably aggravate the fluctuation of loads and the difficulty of the scheduling of an electric vehicle charging station (EVCS) \cite{li2016helos}. Therefore, how to coordinate the dispatching between CIES and EVCS in multi-stakeholder scenarios for reducing the joint operational cost is a challenging issue. 

\subsection{Literature Review}
Currently, there have already been a significant number of investigations on the CIES optimal scheduling. In \cite{wang2018modeling}, the proposed optima strategy improves the economy of the CIES effectively through the coordination of the complementary operation of multiple energy carriers. The inherent volatility and uncertainty of RGs lead to a certain degree of difficulty in the economic scheduling while providing a variety of load supply. By coordination of energy storage systems (ESS) and demand response, the work in \cite{8731714} obtains a higher profit by higher participation of wind power. Furthermore, in terms of demand response, Ref. \cite{8052516} presents a stochastic district optimization model for an integrated energy community to seek the maximization of flexibility potential. Ref. \cite{nguyen2016dynamic} studies the energy scheduling problem of inflexible and flexible loads. However, much attention has been given to single electricity demand response in previous works, and relatively less attention to integrated demand response (IDR). 

With the popularity of EVs, the increasing penetration of EVs has made a profound influence on the optimal scheduling and energy management of CIES. Ref. \cite{taibi2018strategies} considers the additional value brought by EVs to provide energy and auxiliary services to the grid. As EVs have the dual attributes of controllable loads and energy storage equipments \cite{mwasilu2014electric}, guiding them to participate in demand response can not only smooth the source-end fluctuations caused by the high-penetration RGs' uncertainties but also enhance the accommodation rate of RGs \cite{7128736}.

At present, the existing references have researched the coordinated scheduling problem of EVCS. Ref. \cite{7182331} develops an integrated EV charging navigation framework, which  attracts EVs users to charge during off-peak periods. However, only the charging behavior of EVs is considered in the method. To give full play to the  demand response potentials of EVs, \cite{li2016helos} adopts dynamic electricity prices to guide the charging and discharging behaviors of EVs, but it ignores the auxiliary function of flexible EVs in supplying spinning reserve services. Ref. \cite{yu2015real} establishes a model for minimizing energy cost of a residential household with EV, ESS, but the results only consider the benefits of the EV owners. Ref. \cite{wang2015integrated} proposes a price-based decentralized electric vehicle scheduling strategy. The vehicle-to-grid (V2G) optimization method which incorporates EV models into power grid optimization is proposed in \cite{gao2014integrated}. These stakeholders typically act separately during operation and their energy trading problem has not yet been well investigated. In summary, it is of great significance to guide EVs to participate in flexible demand response so that EVs can track RGs' outputs and promote their consumption.

In dealing with the problem of coordinating the interests of multiple stakeholders, bi-level programming is a widely used optimization method. Ref. \cite{talari2017optimal} applies stochastic bilevel programming techniques for solving two objective functions for DR scheduling in a pre-emptive market. In \cite{gu2020bi}, a bi-level optimal low-carbon economic dispatch model is proposed for an industrial park considering multi-energy price incentives. To summarize the unique features of the proposed method with respect to the previous works in the area, the comparison of the proposed method and related work can be seen in Table \ref{tab}. Unfortunately, to the authors' best knowledge there are very few or no references have conducted a comprehensive study on the coordination dispatching between CIES and EVCS in a multi-stakeholder scenario up to now.

\begin{table}[]
  \centering
  \caption{ Comparison of the proposed approach with related works} \label{tab}
  \setlength{\tabcolsep}{0.4mm}{
  \begin{tabular}{c|c|c|c|c|c|c|c}
  \hline
  \multirow{2}{*}{Ref} & \multicolumn{2}{c|}{Stakeholders}                                                                                                                 & \multirow{2}{*}{\begin{tabular}[c]{@{}c@{}} Scheduling modeling \\ method\end{tabular}}    & \multirow{2}{*}{IDR} & \multirow{2}{*}{EV} & \multicolumn{2}{c}{\begin{tabular}[c]{@{}c@{}}Renewable \\ Uncertainties\end{tabular}} \\ \cline{2-3} \cline{7-8} 
                       & upper-level                                                                       & lower-level                                                   &                                                                                              &                      &                     & WT                                         & PV                                         \\ \hline
  {\cite{jin2021optimal}}              & \begin{tabular}[c]{@{}c@{}}intergrated \\ community \\ energy system\end{tabular} & consumers                                                     & \begin{tabular}[c]{@{}c@{}}bi-level model \\ predictive control \\ optimization\end{tabular} & ×                    & ×                   & $\surd$                                    & $\surd$                                    \\ \hline
  \cite{li2016helos}              & \begin{tabular}[c]{@{}c@{}}microgrid \\ operators\end{tabular}                    & EVs users                                                     & \begin{tabular}[c]{@{}c@{}}Lyapunov\\ optimization\end{tabular}                              & ×                    & $\surd$             & ×                                          & ×                                          \\ \hline
  \cite{8731714}              & \multicolumn{2}{c|}{\begin{tabular}[c]{@{}c@{}}single stakeholder \\ i.e. virtual power plant\end{tabular}}                                       & \begin{tabular}[c]{@{}c@{}}multi-stage\\ stochastic\\programming\end{tabular}                             & ×                    & ×                   & $\surd$                                    & ×                                          \\ \hline
  \cite{nguyen2016dynamic}             & \begin{tabular}[c]{@{}c@{}}load serving \\ entity\end{tabular}                    & DR aggreator                                                  & \begin{tabular}[c]{@{}c@{}}bi-level\\ programming\end{tabular}                               & ×                    & ×                   & ×                                          & ×                                          \\ \hline
  \cite{talari2017optimal}             & \begin{tabular}[c]{@{}c@{}}independent \\ system \\ operator\end{tabular}         & DR aggregator                                                 & \begin{tabular}[c]{@{}c@{}}bi-level\\ stochasitic\\ programming\end{tabular}                 & ×                    & ×                   & $\surd$                                    & ×                                          \\ \hline
  \cite{gu2020bi}             & \begin{tabular}[c]{@{}c@{}}integrated \\ energy service\\ agency\end{tabular}    & \begin{tabular}[c]{@{}c@{}}multi-energy \\ users\end{tabular} & \begin{tabular}[c]{@{}c@{}}bi-level\\ optimization\end{tabular}                              & $\surd$              & ×                   & ×                                          & ×                                          \\ \hline
  \cite{li2021optimal}             & \begin{tabular}[c]{@{}c@{}}integrated \\ energy \\ operator\end{tabular}            & users                                                         & \begin{tabular}[c]{@{}c@{}}leader-follower\\ stackelberg\\ game\end{tabular}                 & $\surd$              & ×                   & $\surd$                                    & $\surd$                                    \\ \hline
  \textbf{This paper}  & \textbf{CIES}                                                                     & \textbf{EVCS}                                                 & \textbf{\begin{tabular}[c]{@{}c@{}}bi-level\\ iterative \\ method\end{tabular}}              & $\surd$              & $\surd$             & $\surd$                                    & $\surd$                                    \\ \hline
  \end{tabular}
}
\end{table}

\subsection{Contribution of This Paper}
This study presents a new bi-level optimal dispatching model for the CIES with EVCS in multi-stakeholder scenarios. The contributions of this work are summarized as follows:
\begin{enumerate}

    \item To coordinate flexible demand responses and multiple renewable generations uncertainties, a novel bi-level optimal dispatching model for the CIES with an EVCS is established in this paper, where three different operating modes of EVs (charging, discharging and providing spinning reserves) are fully explored.
    \item  To promote a balance between energy supply and demand, a new integrated demand response program considering flexible thermal comfort requirements of  users is designed by introducing a predictive mean voting (PMV) index. By this means, the electricity and heating demands of the CIES are met while maintaining a user comprehensive satisfaction within an acceptable range.
    \item To further explore the potential of demand response, a dynamic pricing mechanism that combines time-of-use (TOU) and real-time (RT) pricing is proposed. This mechanism can flexibly guide users' energy consumption  and EVs charging/discharging behaviors to consume renewable energy.
     \item A simulation test was performed on a CIES located in North China to verify the effectiveness and superiority of the proposed method. The influences of main control parameters on the performance of the proposed method have also been analyzed in detail.
\end{enumerate}

\section{Modeling of CIES}\label{The model}
\subsection{Renewable Generations Models}
Since wind turbines (WT) and photovoltaic (PV) power outputs are both random variables, probability density functions (PDF) are usually used to describe their uncertainties. Among them, WT and PV outputs respectively obey the Beta distribution and the Weibull distribution. Ref. \cite{li2018optimal1} gives a detailed description of their respective PDFs and the corresponding derivation process.
\subsection{Integrated Demand Response}
\subsubsection{Electricity Demand Response}
Electric load consists of fixed load and flexible load in this study. According to the characteristics of demand response, the electric flexible loads is divided into two types: shiftable load and interruptible load. 
\paragraph{Shiftable electrical load}
The characteristic of the shiftable load is that the total amount of electricity consumption is constant, and the consumption time can be flexibly changed \cite{ma2016residential}. It can be described by the following formula:
{\setlength\abovedisplayskip{2pt}
\setlength\belowdisplayskip{2pt}
\begin{equation}\label{TSL1}
    P_{{t,\min }}^{TSL}\le P_{t}^{TSL}\le P_{t,\max }^{TSL}
\end{equation} }{\setlength\abovedisplayskip{2pt}
\setlength\belowdisplayskip{2pt}
\begin{equation}\label{TSL2}
   \sum\limits_{i=1}^{T}{P_{t}^{TSL}}\textit{=}0
\end{equation} }where $P_{t}^{TSL}$ is the shifted electrical load power, $P_{{t,\max }}^{TSL}$ and $P_{{t,\min }}^{TSL}$ are the maximum and minimum values of the shiftable load during period $t$. 
\paragraph{Interruptible electrical load}
During periods of insufficient power supply or high electricity price, users can interrupt part of the load to relieve the pressure of the power supply \cite{wang2020economic}. The related constraints of interruptible load can be described as: 
{\setlength\abovedisplayskip{2pt}
\setlength\belowdisplayskip{2pt}
\begin{equation}
   0\le P_{t}^{EIL}\le P_{t,\max }^{EIL}
\end{equation} }where $P_{t}^{EIL}$ and $P_{t,\max }^{EIL}$ are the interrupted electric power and its maximum value in period $t$ respectively. The maximum interruptible load power is 10\% of the electric load demand in each period.

\subsubsection{Heating Demand Response}

This paper considers the building heat demand as the heat load, and the transient heat balance equation can be utilized to link the building temperature with the heat demand
\cite{li2021optimal}.

Changing the indoor temperature within a certain range will not affect the user's comfort experience, that is, the user's perception of the heating temperature has a certain degree of ambiguity. Here, the thermal sensation vote value is used to describe the user's comfortable experience of indoor temperature changes. The related constraint of heating interruptible load can be described as: 
{\setlength\abovedisplayskip{2pt}
\setlength\belowdisplayskip{2pt}
\begin{equation}
   0\le P_{t}^{HIL}\le P_{t,\max }^{HIL}
\end{equation} }where $P_{t}^{HIL}$ and $P_{t,\max }^{HIL}$ are the interrupted heating power and its maximum value in period $t$, respectively.
The Predictive Mean Vote (PMV) index is introduced to describe the acceptable thermal comfort range of users \cite{li2021optimal}:
{\setlength\abovedisplayskip{2pt}
\setlength\belowdisplayskip{2pt}
\begin{equation}\label{LS}
  PMV=2.43-\frac{3.76\left( {{T}_{s}}-{{T}_{in}} \right)}{M\left( {{I}_{cl}}+0.1 \right)}
\end{equation} }where $M$ is the human energy metabolism rate; $I_{cl}$ is the thermal resistance of clothing; ${T}_{s}$ is the average temperature of the human skin in a comfortable state; ${T}_{in}$ is the indoor temperature. The indoor temperature change range is

{\setlength\abovedisplayskip{2pt}
\setlength\belowdisplayskip{2pt}
\begin{small}
\begin{equation}
  \left\{ \begin{aligned}
  & \left| PMV \right|\le 0.9,\text{   }t\in \text{ }\!\![\!\!\text{ 1:00-7:00 }\!\!]\!\!\text{ }\cup \text{ }\!\![\!\!\text{ 20:00-24:00 }\!\!]\!\!\text{ } \\ 
 & \left| PMV \right|\le 0.5,\text{   }t\in \text{ }\!\![\!\!\text{ 8:00-19:00 }\!\!]\!\!\text{ } \\ 
\end{aligned} \right.
\end{equation} 
\end{small} }
To comprehensively measure the impact of IDR on users' experience, inspired by the work in \cite{li2019power}, a user comprehensive satisfaction is designed as
{\setlength\abovedisplayskip{2pt}
\setlength\belowdisplayskip{2pt}
\begin{equation}\label{LS}
 US=\frac{\left( \frac{P_{t}^{GL}+P_{t}^{TSL}-P_{t}^{EIL}}{P_{t}^{EL}} \right)+\left( \frac{P_{t}^{HL}-P_{t}^{HIL}}{P_{t}^{HL}} \right)}{2}\times 100\%
\end{equation} }where $US$ is the user  comprehensive  satisfaction, $P_{t}^{GL}$ is the uninterruptible load in period $t$, and $P_{t}^{EL}$ and $P_{t}^{HL}$ are the initial electric and heat loads. 

\subsection{Electric Vehicle Charging Station}

The U.S. private car driving survey in 2009 shows that the time of EVs arriving the EVCS approximately obeys a normal distribution \cite{Federal}. The probability density function (PDF) of the arrival time of EVs is described by (\ref{FEV1}):

{\setlength\abovedisplayskip{2pt}
\setlength\belowdisplayskip{2pt}
\begin{footnotesize}
\begin{equation}\label{FEV1}
{{f}^{ar}}_{EV}(t)=\left\{ \begin{aligned}
  & \frac{1}{\sqrt{2\pi }{{\sigma }_{1}}}\exp [-\frac{{{(t+24-{{\mu }_{1}})}^{2}}}{2\sigma _{1}^{2}}]\text{   }0<t\le {{\mu }_{1}}-12 \\ 
 & \frac{1}{\sqrt{2\pi }{{\sigma }_{1}}}\exp [-\frac{{{(t-{{\mu }_{1}})}^{2}}}{2\sigma _{1}^{2}}]\quad\quad\text{           }{{\mu }_{1}}-12<t\le 24 \\
\end{aligned} \right.
\end{equation}
\end{footnotesize} }where ${{\mu }_{1}}$ and ${{\mu }_{2}}$ are the mean values of the time when EVs arrive and depart from EVCS, respectively; ${{\sigma }_{1}}$ and ${{\sigma }_{2}}$ are the standard deviations of the time when EVs arrive and leave EVCS, respectively.

The daily load demand of EVs charging is related to the daily driving mileage and charging duration. In general, the daily travel mile of an EV is considered to obey a normal distribution, and its PDF is:
{\setlength\abovedisplayskip{2pt}
\setlength\belowdisplayskip{2pt}
\begin{equation}\label{LS}
{{f}_{M}}({{M}_{d}})=\frac{1}{\sqrt{2\pi }{{\sigma }_{M}}{{M}_{d}}}\exp [-\frac{{{(\ln {{M}_{d}}-{{\mu }_{M}})}^{2}}}{2\sigma _{M}^{2}}]
\end{equation} }where ${{M}_{d}}$ represents the daily mileage of EVs; ${{\sigma }_{M}}$ and ${{\mu }_{M}}$ are the standard deviation and the mean value of ${{M}_{d}}$.

Based on the travel mileage of an EV and its initial state of charge (SOC), the actual SOC at the end of the charging is {\setlength\abovedisplayskip{2pt}
\setlength\belowdisplayskip{2pt}
\begin{equation}\label{LS}
{{S}_{real}}={{S}_{s}}+\frac{{{M}_{d}}{{E}_{d,100}}}{100{{B}_{c}}}\end{equation} }where ${{S}_{real}}$ denotes the real charging state, ${{S}_{s}}$ is the initial SOC of the EV, ${{E}_{d,100}}$ is the power demand when the EV travels 100 kilometers, ${{B}_{c}}$ indicates the battery capacity of the EV.
The charging time of an EV can be calculated by
{\setlength\abovedisplayskip{2pt}
\setlength\belowdisplayskip{2pt}
\begin{equation}\label{charging time}{{T}_{ch}}=\frac{(S_{real}^{{}}-S_{s}^{{}})C_{t}^{EV}}{P_{rated}^{EV}\eta _{ch}^{EV}}\end{equation} }where  ${{T}_{ch}}$ is the charging time of the EV; $P_{rated}^{EV}$ and $\eta _{ch}^{EV}$ are the rated electricity power and the charging efficiency of the EV; $C_{t}^{EV}$ is the battery capacity of the EV.

According to (\ref{FEV1})-(\ref{charging time}), Monte Carlo simulation (MCS) is used to simulate the charging demand of EVs in a disordered state, and then the obtained  EV disordered charging demand is incorporated in the bi-level model as the initial charge-discharge scheme of the EVCS in the first iteration process.

\section{Problem formulation}\label{Problem formulation}
To coordinate the dispatching problems between CIES and EVCS, this paper proposes a novel bi-level dispatching model, in which the upper and lower-level are formulated to minimize the operating costs of the CIES and EVCS, respectively. A dynamic pricing mechanism “TOU+RT” is proposed as a bridge between the two levels.
\subsection{Dynamic Pricing Mechanism}
Designing a pricing mechanism based on supply and demand has been proven to be highly effective by existing researches \cite{jin2021optimal},\cite{li2016helos},\cite{8731714},\cite{nguyen2016dynamic},\cite{li2018optimal1}. Based on this principle, the dynamic pricing mechanism proposed fully combines the advantages of TOU prices and RT prices, which can efficiently guide the charging-discharging scheme of lower-level EVs users, and effectively reduce upper and lower-level operating costs. The proposed pricing mechanism can be described as

{\setlength\abovedisplayskip{2pt}
\setlength\belowdisplayskip{2pt}
\begin{footnotesize}
\begin{equation}
\begin{aligned}\label{LS}
 {{P}_{load,t}} &= \left( P_{t}^{TSL}+P_{t}^{GL}-P_{t}^{IL}-P_{DC,t}^{ESS}+P_{CH,t}^{ESS} \right) \\
& +\left( P_{t}^{HL}-P_{t}^{HIL}-P_{DC,t}^{HSS}+P_{CH,t}^{HSS} \right)-\left( P_{DC,t}^{EV}-P_{CH,t}^{EV} \right)
\end{aligned}
\end{equation} 
\end{footnotesize}}
{\setlength\abovedisplayskip{2pt}
\setlength\belowdisplayskip{2pt}
\begin{equation}\label{LS}
{{\omega }_{rt,t}}=k({{P}_{load,t}},E(P_{t}^{RGs}))\cdot {{\omega }_{st.t}}
\end{equation} }
{\setlength\abovedisplayskip{2pt}
\setlength\belowdisplayskip{2pt}
\begin{footnotesize}
\begin{equation}\label{LS1}
k({{P}_{load,t}},E(P_{t}^{RGs}))\text{=}\left\{ \begin{aligned}
  & 1 \text{, }{{P}_{load,t}}>E(P_{t}^{RGs}) \\ 
 & {{P}_{load,t}}/E(P_{t}^{RGs})\text{, }{{P}_{load,t}}\le E(P_{t}^{RGs})\text{ } \\ 
\end{aligned} \right.
\end{equation} 
\end{footnotesize}}
{\setlength\abovedisplayskip{2pt}
\setlength\belowdisplayskip{2pt}
\begin{footnotesize}
\begin{equation}\label{LS2}
{{\omega }_{rt,t}}\text{=}\left\{ \begin{aligned}
  & {{\omega }_{\text{s}t,t}} \text{ , }{{P}_{load,t}}>E(P_{t}^{RGs}) \\ 
 & k({{P}_{load,t}},E(P_{t}^{RGs}))\cdot {{\omega }_{\text{s}t}^{*}}\text{, }  {{P}_{load,t}}\le E(P_{t}^{RGs})\text{ } \\ 
\end{aligned} \right.
\end{equation} 
\end{footnotesize}}where ${P}_{load,t}$ represents the total load demand of the CIES; $P_{CH,t}^{ESS}$ and $P_{DC,t}^{ESS}$ is the charge and disharge powers of ESS; $P_{CH,t}^{HSD}$ and $P_{DC,t}^{HSD}$ are heat storage and release powers of the heat storage device (HSD); $P_{CH,t}^{EV}$ and $P_{DC,t}^{EV}$ are the EV charge and discharge powers; $P_{t}^{RGs}$ and $E(P_{t}^{RGs})$ are the joint outputs of RGs and its expected value; $\displaystyle k({{P}_{load,t}},E(P_{t}^{RG})$ represents supply and demand relationship of the CIES; ${{\omega }_{rt,t}}$ is the dynamic prices passed from the CIES to the EVCS; ${{\omega }_{st,t}}$ is the grid TOU prices; and ${\omega }_{st}^{*}$ is the gird TOU prices in valley periods. 

Through the dynamic pricing mechanism, the CIES provides price concessions to users, so users can choose the periods with lower electricity prices to reduce the electricity cost. Therefore, users have enough motivations to adjust electricity consumption behaviors and actively participate in the IDR.

The overall pricing process is as follows: (1) Firstly, obtain $E(P_{t}^{RGs})$ by using sequence operation theory (SOT). (2) After that, the charging-discharging scheme of EVs users is obtained through the optimization of the lower-level model and fed back to the upper level. (3) Finally, the upper-level obtains ${{\omega }_{rt,t}}$ through the dynamic pricing mechanism.
\subsection{The Upper-level Model}
\subsubsection{Objective Function}

The upper-level model takes the minimization of the CIES net operating cost as the objective function. The objective function is specifically described as follows:

{\setlength\abovedisplayskip{2pt}
\setlength\belowdisplayskip{2pt}
\begin{footnotesize}
\begin{equation}
\begin{aligned}\label{F1}
  & \min {{F}_{1}}=\sum\limits_{t=1}^{T}{{{\omega }_{st,t}}\left( P_{e,t}^{TSL}+P_{e,t}^{GL}-P_{e,t}^{IL}+P_{e,t}^{HL} \right)}+\sum\limits_{t=1}^{T}{{{\omega }_{rt,t}}P_{DC,t}^{EV}} \\ 
 & +\sum\limits_{t=1}^{T}{\left( {{\omega }_{re,grid}}R_{t}^{grid}+{{\omega }_{re,ESS}}R_{t}^{ESS}+{{\omega }_{re,EV}}R_{t}^{EV} \right)} \\ 
 & +\sum\limits_{t=1}^{T}{{{\omega }_{dp,ESS}}P_{CH,t}^{ESS}}+\sum\limits_{t=1}^{T}{{{\omega }_{el}}P_{e,t}^{EIL}+ {\omega }_{hl}}P_{e,t}^{HIL}-\sum\limits_{t=1}^{T}{{{\omega }_{rt,t}}P_{RGs,t}^{EV}}
\end{aligned}
\end{equation}
\end{footnotesize}}where, in period $t$, $P_{e,t}^{TSL}$, $P_{e,t}^{GL}$ and $ P_{e,t}^{HL}$ are the grid power consumed by the shiftable electrical load, unshiftable electrical load and heat load, respectively. ${{\omega }_{re,grid}}$, $ {{\omega }_{re,ESS}}$ and ${{\omega }_{re,EV}}$ are the grid, ESS and EVs reserve prices, respectively. $R_{t}^{grid}$, $R_{t}^{ESS}$ and $R_{t}^{EV}$ are the reserve capacities provided by the grid, ESS and EVs, respectively. $ {{\omega }_{dp,ESS}}$ is the depreciation cost of ESS, ${{\omega }_{el}}$ and ${{\omega }_{hl}}$ are the electric and heating compensation prices for interruption loads, respectively; and $ P_{RGs,t}^{EV}$ is the RGs power consumed by EVs.

\subsubsection{Constraint Conditions}
\paragraph{Power supply system constraints}
Eq. (\ref{eq13}) reflects the balance of power supply and demand of CIES. Eq. (\ref{eq15}) is the constraint of the power provided by the grid.

{\setlength\abovedisplayskip{2pt}
\setlength\belowdisplayskip{2pt}
\begin{small}
\begin{equation}\label{eq13}
\begin{aligned}
   P_{EL,t}^{grid}+&E(P_{t}^{RGs})+P_{DC,t}^{EV}+P_{DC,t}^{ESS}-P_{CH,t}^{ESS} \\ 
 & \text{                       =}P_{t}^{TSL}+P_{t}^{GL}-P_{t}^{IL}+P_{CH,t}^{EV}+P_{t}^{CL},\forall t \\ 
\end{aligned}
\end{equation}
\end{small} }
\begin{equation}\label{eq15}
P_{t}^{grid}+R_{t}^{grid}\le P_{\text{max}}^{grid},\forall t
\end{equation}where, $P_{EL,t}^{grid}$ is the grid power consumed by the system power load in period $t$, and $P_{t}^{CL}$ is the controllable load, and $P_{\max }^{grid}$ is the maximum power provided by the grid.
\paragraph{ESS constraints}
Eqs. (\ref{eq17})-(\ref{eq19}) give the operational constraints of ESS capacity and charge/discharge power. Eq. (\ref{eq20}) ensures that the ESS operation in each cycle has the same initial state. 
{\setlength\abovedisplayskip{2pt}
\setlength\belowdisplayskip{2pt}
\begin{equation}\label{eq17}
C_{t+1}^{ESS}=C_{t}^{ESS}+({{\eta }_{ch}}P_{CH,t}^{ESS}-P_{DC,t}^{ESS}/{{\eta }_{dc}})\Delta t,\forall t
\end{equation}
}
\begin{equation}\label{eq18}
\left\{ \begin{aligned}
  & 0\le P_{DC,t}^{ESS}\le P_{DC,\text{max}}^{ESS} \\ 
 & 0\le P_{CH,t}^{ESS}\le P_{CH,\text{max}}^{ESS} \\ 
\end{aligned} \right.\forall t
\end{equation}
\begin{equation}\label{eq19}
C_{\min }^{ESS}\le C_{t}^{ESS}\le C_{\max }^{ESS},\forall t
\end{equation}
\begin{equation}\label{eq20}
C_{0}^{ESS}=C_{{{\text{T}}_{\text{end}}}}^{ESS}=C_{_{\text{*}}}^{ESS}
\end{equation}
where $C_{t}^{ESS}$ is the ESS capacity, $\eta _{ch}^{{}}$ and $\eta _{dc}^{{}}$ are the ESS charging and discharging efficiencies. $C_{0}^{ESS}$ is the initial capacity of the ESS, $C_{{{T}_{end}}}^{ESS}$ is the ESS capacity at the end of a dispatching period (set to 24 hours), and $C_{\text{*}}^{ESS}$ is the minimum initial ESS capacity.

The reserve capacity provided by the ESS is supposed to meet the following constraint:

\begin{footnotesize}
\begin{equation}
{{P}_{Ress,t}}\le \min \left\{ {{\eta }_{dc}}(C_{t}^{ESS}-C_{\text{min}}^{ESS})/\Delta t,P_{DC,\max }^{ESS}-P_{DC,t}^{ESS} \right\},\forall t
\end{equation}
\end{footnotesize}where ${{P}_{Ress,t}}$ is the reserve capacity supplied by ESS.

Besides, the spinning reserve of the entire system is provided by the grid, ESS and EVs, which is described as the following chance constraint:

{\setlength\abovedisplayskip{2pt}
\setlength\belowdisplayskip{2pt}
\begin{scriptsize}
\begin{equation}\label{eq22}
\begin{aligned}
  {{P}_{rob}}\Bigg\{{R_{t}^{grid}}+{{P}_{Ress,t}}+R_{t}^{EV}\ge 
   E(P_{t}^{RGs})- 
   P_{t}^{WT}-P_{t}^{PV} \Bigg\}\ge \alpha ,\forall t 
\end{aligned}
\end{equation}
\end{scriptsize}}where $\alpha $ is the preset confidence level of the spinning reserve constraint.
\paragraph{Heating system constraints}

Electrical and heat power balance constraints:
{\setlength\abovedisplayskip{2pt}
\setlength\belowdisplayskip{2pt}
\begin{small}
\begin{equation}
 \sum\limits_{n=1}^{N}{{{P}_{dhp,n,t}}}=P_{RGs,t}^{HL}+P_{e,t}^{HL}=P_{t}^{HL} 
\end{equation}
\end{small}}where ${{P}_{dhp,n,t}}$ is the electricity consumption of heating load, $ P_{e,t}^{HL}$ is the RGs power consumed by the heat load, ${{P}_{eh,n,t}}$ is the heating power of the $n$th electric boiler (EB), $N$ is the total number of EBs.

Electric boiler constraints:
{\setlength\abovedisplayskip{2pt}
\setlength\belowdisplayskip{2pt}
\begin{equation}\label{eqEB1}
{{P}_{eh,n,t}}=\eta_{eb} {{P}_{dhp,n,t}}
\end{equation}
\begin{equation}\label{eqEB2}
0\le {{P}_{eh,n,t}}\le {{P}_{eh,n}}
\end{equation}}where ${{P}_{eh,n,t}}$ is the rated heating power of the EB, $\eta_{eb}$ is the EB performance factor, which represents the ratio of heat pump heating power to power consumption.

Heat storage device constraints:
The constraints of the power and capacity of storing/releasing heat that HSD obeys are similar to those of the ESS, namely (\ref{eq17})-(\ref{eq20}). 

\subsection{The Lower-level Model}
EVs users can reduce their own charging costs through discharging behavior, while also alleviating the pressure on power supply of system. In addition, EVs not only make their users compensated by providing spinning reserve services, but also maintain the secure and reliable operation of the CIES \cite{bai2015robust}. Given this, the lower objective function is constructed.
\subsubsection{Objective Function}
The lower-level model takes the minimization of the EVCS net operating cost as the objective function, which is
{\setlength\abovedisplayskip{2pt}
\setlength\belowdisplayskip{2pt}
\begin{equation}\label{eq31}
\begin{aligned}
 & \min {{F}_{2}} ={{\omega }_{st,t}}P_{grid,t}^{EV}+{{\omega }_{rt,t}}\left( P_{RGs,t}^{EV}-P_{DC,t}^{EV} \right) \\
 & -{{\omega }_{re,EV}}R_{t}^{EV}  
\end{aligned}
\end{equation}}where $P_{grid,t}^{EV}$ is the grid power consumed by EVs.
\subsubsection{Constraint Conditions}
\paragraph{EVs power balance constraints}
EVs charge and discharge power do not exceed the allowable range while maintaining the balance of system power. Meanwhile, the discharge power should not exceed the total power deficits required by electric and heat loads. These constraints are expressed as follows:
{\setlength\abovedisplayskip{2pt}
\setlength\belowdisplayskip{2pt}
\begin{equation}
\left\{ \begin{aligned}
  & 0\le P_{DC,t}^{EV}\le \min \left( P_{DC,\text{max}}^{EV},P_{EL,t}^{ed}+P_{HL,t}^{ed} \right) \\ 
 & 0\le P_{CH,t}^{EV}\le P_{CH,\text{max}}^{EV} \\ 
\end{aligned} \right.\forall t
\end{equation} }where, $P_{CH,\text{max}}^{EV}$ and $P_{DC,\max }^{EV}$ are the maximum charging and discharge power of EVs in period $t$, $P_{EL,t}^{de}$ and $P_{HL,t}^{de}$ are the power deficits of the electric load and heat load in period $t$ after consuming RGs power. 

To keep the power balance, the spinning reserve capacity cannot exceed the reserve capacity that the grid should have provided to CIES. At the same time, the reserve capacity of EVs does not exceed its available capacity. This constraint is
{\setlength\abovedisplayskip{2pt}
\setlength\belowdisplayskip{2pt}
\begin{equation}
R_{t}^{EV}\le \min \left\{ P_{DC,\max }^{EV}-P_{DC,t}^{EV},R_{t}^{ed} \right\},\forall t
\end{equation} }where $R_{t}^{de}$ is the additional capacity required by the upper-level when the ESS reserve capacity is insufficient.

The power purchased by EVs from CIES cannot exceed the controllable load power, that is, it cannot exceed the surplus of RGs power in the CIES, which is expressed as
{\setlength\abovedisplayskip{2pt}
\setlength\belowdisplayskip{2pt}
\begin{equation}
0\le P_{RG,t}^{EV}\le P_{t}^{CL}
\end{equation} }\;\; The power purchased by EVs from the grid cannot exceed the upper limit of the power provided by the grid, which is formulated as
{\setlength\abovedisplayskip{2pt}
\setlength\belowdisplayskip{2pt}
\begin{equation}
0\le P_{grid,t}^{EV}\le P_{\max ,t}^{grid}
\end{equation} }\;\;The total charging power of EVs cannot exceed the allowable range in a dispatching period, which is described as
{\setlength\abovedisplayskip{2pt}
\setlength\belowdisplayskip{2pt}
\begin{equation}
P_{\min }^{EV}<P_{t}^{EV}<P_{\max }^{EV}
\end{equation} }
\paragraph{EVs battery constraints:}
Eqs. (\ref{eq37})-(\ref{eq39}) indicates that the total capacity of the EVs battery and the number of charge and discharge cells must be within the allowable range:
{\setlength\abovedisplayskip{2pt}
\setlength\belowdisplayskip{2pt}
\begin{equation}\label{eq37}
C_{t+1}^{EV}=C_{t}^{EV}+(\eta _{ch}^{EV}P_{CH,t}^{EV}-P_{DC,t}^{EV}/\eta _{dc}^{EV})\Delta t,\forall t
\end{equation} }
\begin{equation}\label{eq38}
C_{\min }^{EV}\le C_{t}^{EV}\le C_{\max }^{EV},\forall t
\end{equation}
\begin{equation}\label{eq39}
\left\{ \begin{aligned}
  & N_{CH,t}^{EV}\le {{N}_{B,pos,\max }} \\ 
 & N_{DC,t}^{EV}\le {{N}_{B,pos,\max }} \\ 
\end{aligned} \right.
\end{equation}
where $\eta _{dc}^{EV}$ is the discharging efficiencies of the EVs.
$C_{\min }^{EV}$ and $C_{\max }^{EV}$ are the minimum and maximum capacities of EVs in period $t$, respectively.
 $N_{CH,t}^{EV}$ and $N_{DC,t}^{EV}$ are the number of EVs that are charging/discharging powers.
\section{Model solving}\label{Model solving}

\subsection{Probabilistic Serialization Description of RGs}
The probabilistic sequence can be expressed as \cite{wang2016dependent}:
{\setlength\abovedisplayskip{2pt}
\setlength\belowdisplayskip{2pt}
\begin{equation}\label{eq40}
\sum\limits_{i=0}^{{{N}_{a}}}{a(i)=1},a(i)\ge 0
\end{equation}
}

Taking WT generation as an example, since its PDF is known, a special discretization process is performed on the continuous probability distribution to obtain the corresponding probability sequence. The probabilistic sequence length is
{\setlength\abovedisplayskip{2pt}
\setlength\belowdisplayskip{2pt}
\begin{equation}\label{eq41}
{{N}_{a,t}}=[P_{\max ,t}^{WT}/q]
\end{equation} }where $q$ is the discretization step size; $P_{\max ,t}^{WT}$ is the maximum possible wind power. The probabilistic sequence obtained by discretization has a total of ${{N}_{(a,t)}}$ states, of which the output of the ${{u}_{a}}$ state is ${{u}_{a}}q$, and the corresponding probability is $a({{u}_{a}})$. Table \ref{tab1} lists the power of the WT and the corresponding probability sequence.

\begin{table}[]
\centering
\caption{WT output and its corresponding probabilistic sequence}\label{tab1}
\begin{tabular}{@{}ccccccc@{}}
\toprule
Power/kW & 0 & $q$ & $…$ & $U_aq$  & $…$  & $N_{a,t}q$  \\\midrule
Probability & $a(0)$ & $a(1)$ & $…$ & $a(u_a)$ & $…$ & $a(N_{a,t})$\\ \bottomrule
\end{tabular}
\end{table}

According to the PDF of WT output $f_0(P^{WT})$, the corresponding probability sequence can be obtained, and the calculation formula is shown in (\ref{eq42}).

{\setlength\abovedisplayskip{2pt}
\setlength\belowdisplayskip{2pt}
\begin{footnotesize}
\begin{equation}\label{eq42}
a({{i}_{a,t}})=\left\{ \begin{aligned}
  & \int_{0}^{q/2}{{{f}_{o}}(P_{{}}^{WT})dP_{{}}^{WT},{{i}_{a,t}}=0} \\ 
 & \int_{{{i}_{a,t}}q-q/2}^{{{i}_{a,t}}q+q/2}{{{f}_{o}}(P_{{}}^{WT})dP_{{}}^{WT},{{i}_{a,t}}>0,{{i}_{a,t}}\ne {{N}_{a,t}}} \\ 
 & \int_{{{i}_{a,t}}q-q/2}^{{{i}_{a,t}}q}{{{f}_{o}}(P_{{}}^{WT})dP_{{}}^{WT},{{i}_{a,t}}={{N}_{a,t}}} \\ 
\end{aligned} \right.
\end{equation}
\end{footnotesize} }

Similarly, the same method can be used to handle the probability sequence corresponding to the WT output.

When using the SOT, the proper selection of the step size $q$ plays an important role in optimal results. A smaller step will lead to accurate, but a low-efficiency calculation; while a larger step can save calculation time but it makes the generated sequences unable to fully reflect the actual probability distributions. To this end, a sensitivity analysis of the step size $q$ is performed to achieve a compromise between reliability and economy of the CIES in this study.
\subsection{Handling of Chance Constraints}
In period $t$, the probabilistic sequence $c({{i}_{\text{c},t}})$ corresponding to the joint outputs of RGs is obtained by the addition-type-convolution of the probabilistic sequences $a({{i}_{\text{a},t}})$ and $b(i{}_{\text{b},t})$:
{\setlength\abovedisplayskip{2pt}
\setlength\belowdisplayskip{2pt}
\begin{equation}\label{eq43}
\begin{aligned}
  & c({{i}_{\text{c},t}})=a({{i}_{\text{a},t}})\oplus b(i{}_{\text{b},t})= \\ 
 & \sum\limits_{{{i}_{\text{a},t}}+{{i}_{\text{b},t}}={{i}_{\text{c},t}}}{a({{i}_{\text{a},t}})b(i{}_{\text{b},t}),\text{ }{{i}_{\text{c},t}}=0,1,...,{{N}_{\text{a},t}}+{{N}_{\text{b},t}}} \\
\end{aligned}
\end{equation} }

To deal with (\ref{eq22}), a 0-1 variable ${{W}_{{{u}_{e,t}}}}$ is introduced as
{\setlength\abovedisplayskip{2pt}
\setlength\belowdisplayskip{2pt}
\begin{footnotesize}
\begin{equation}\label{eq44}
{{W}_{{{u}_{e,t}}}}=\left\{ \begin{aligned}
  & 1,\text{}  R_{t}^{gird}+{{P}_{Ress,t}}+R_{t}^{EV}\ge  \\ 
 & \text{           }E(P_{t}^{RGs})-P_{t}^{WT}-P_{t}^{PV} \\
 & 0,otherwise \\ 
\end{aligned} \right.\forall t,{{u}_{e,t}}=0,1,...,{{N}_{e,t}}
\end{equation}
\end{footnotesize} }\; Eq. (\ref{eq44}) suggests that for any period $t$, when the total reserve capacity $R_{t}^{gird}+{{P}_{Ress,t}}+R_{t}^{EV}$ is not less than the margin between the RGs' joint outputs and its expected value, then ${{W}_{{{u}_{e,t}}}}$ is taken as 1; otherwise, it is 0.

As shown in Table \ref{tab2}, each possible joint output of RGs ${{u}_{c,t}}q$ corresponds to a probability of $c({{u}_{c,t}})$. 
\begin{table}[]
\centering
\caption{Probability sequence of RGs' joint outputs}\label{tab2}
\begin{tabular}{@{}ccccccc@{}}
\toprule
 Power/kW & 0 & $q$ & $2q$ & $…$  & $(N_{ c,t }-1)q$  & $N_{c,t}q$  \\\midrule
 Probability & $c(0)$ & $c(1)$ & $c(2)$ & $…$ & $c(N_{c,t}-1)$ & $c(N_{c,t})$\\ \bottomrule
\end{tabular}
\end{table}
Therefore, (\ref{eq22}) can be simplified into the following form:
{\setlength\abovedisplayskip{2pt}
\setlength\belowdisplayskip{2pt}
\begin{equation}\label{eq45}
\sum\limits_{{{u}_{e,t}}=0}^{{{N}_{e,t}}}{{{W}_{{{u}_{e,t}}}}e({{u}_{e,t}})}\ge \alpha 
\end{equation} }\quad In order to make (\ref{eq44}) compatible with mixed-integer linear programming (MILP) formulation, it is replaced with (\ref{eq46}):

{\setlength\abovedisplayskip{2pt}
\setlength\belowdisplayskip{2pt}
\begin{footnotesize}
\begin{equation}\label{eq46}
\begin{aligned}
   (R_{t}^{gird}+{{P}_{Ress,t}}+R_{t}^{EV}-E(P_{t}^{RGs})+P_{t}^{WT}+P_{t}^{PV})/\tau \le {{W}_{{{u}_{e,t}}}}
  \le  & \\  1+(R_{t}^{gird}+{{P}_{Ress,t}}+R_{t}^{EV}-E(P_{t}^{RGs})+P_{t}^{WT}+P_{t}^{PV})/\tau, &\\
 \forall t,{{u}_{e,t}}=0,1,...,{{N}_{e,t}} &
\end{aligned}
\end{equation}
\end{footnotesize}}where $\tau $ is a very large positive number, when $R_{t}^{gird}+{{P}_{Ress,t}}+R_{t}^{EV}\ge E(P_{t}^{RGs})-P_{t}^{WT}-P_{t}^{PV}$, (\ref{eq46}) is equivalent to $\lambda \le {{W}_{{{u}_{e,t}}}}\le 1+\lambda$, since ${{W}_{{{u}_{e,t}}}}$ is a binary variable, therefore, ${{W}_{{{u}_{e,t}}}}$ can only be 1 or 0. Replace (\ref{eq22}) with (\ref{eq45}) and (\ref{eq46}), thereby transforming the chance-constrained programming (CCP) into a MILP formulation.
\subsection{Determination of Dispatching Scheme}

In order to achieve the optimal economic benefits, the joint optimization objective function is represented as:
{\setlength\abovedisplayskip{2pt}
\setlength\belowdisplayskip{2pt}
\begin{equation}\label{eq47}
{{F}^{JO}}=\min \left(\sqrt{{{\left( F_{1}^{JO} \right)}^{2}}+{{\left( F_{2}^{JO} \right)}^{2}}} \right)
\end{equation}}where $F_{1}^{JO}$ and $F_{2}^{JO}$ represent the operating costs of the upper (CIES) and lower-level (EVCS) in the joint optimization solution process, respectively.

\subsection{Solution Process}
\begin{figure}[t]
    \centering
    \includegraphics[width=3.4in]{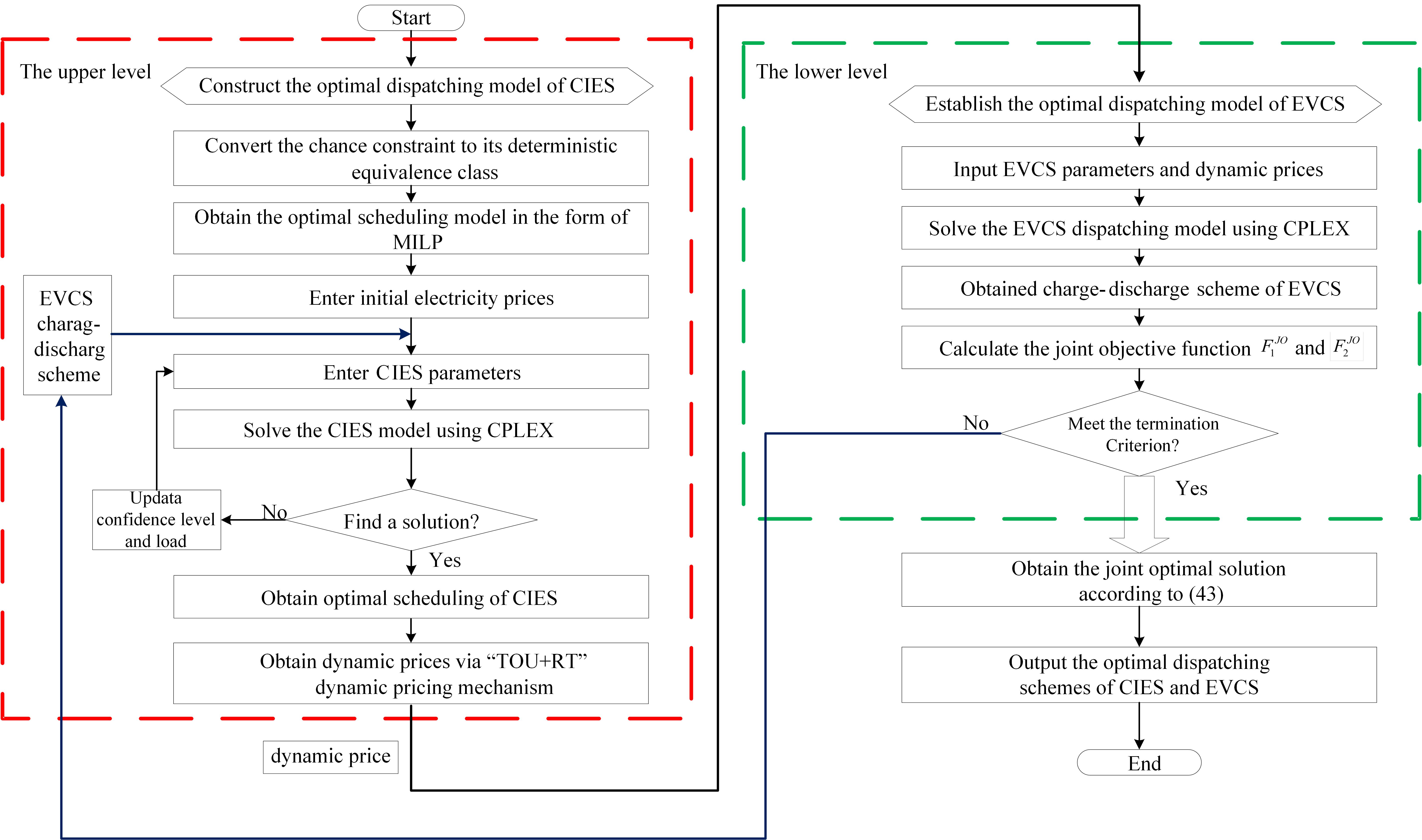}
    \caption{Flowchart of solution process}
    \label{liucheng}
\end{figure}

The solution process of the scheduling model is illustrated in Fig. \ref{liucheng}. The detailed procedures are listed as below:

Step 1: Establish the upper-level CIES optimal dispatching model according to (\ref{F1})-(\ref{eqEB2});

Step 2: Convert the chance constraint to its deterministic equivalence class; 

Step 3: Enter CIES parameters;

Step 4: Determine whether the solution exists. If the solution exists, continue the solution steps; otherwise, update the confidence level and load, and go back to step 3;

Step 5: Obtain the CIES optimal dispatching scheme and dynamic prices through dynamic pricing mechanism, and pass the dynamic prices to the lower-level;

Step 6: Construct the lower-level EVCS optimal dispatching model according to (\ref{eq31})-(\ref{eq39});

Step 7: Enter EVCS parameter;

Step 8: According to dynamic prices provided by the upper-level model,
solve EVCS optimal dispatching model;

Step 9: Obtain the EVCS charge and discharge scheme;

Step 10: Calculate the joint optimization objective function $F_{1}^{JO}$ and $F_{2}^{JO}$ ;

Step 11: Judge whether the termination condition is met. Here, the adopted criterion is the current iteration number exceeds the preset maximum number of iterations. If met, stop the iteration process; otherwise, pass the EVCS charge-discharge scheme to the upper-level and return to step 3;

Step 12: Determine the joint optimal solution by (\ref{eq47});

Step 13: Output the optimal dispatching schemes of the CIES and the EVCS.

\section{Case study}\label{Case study}
To examine the effectiveness of the presented method, a practical example of CIES in winter in North China is used for simulation analysis. All programs are performed on a PC computer with Intel Core I5-5200U CPU and 4G RAM.
\subsection{Parameters Settings}
\begin{figure}[t]
    \centering
    \includegraphics[width=3.4in]{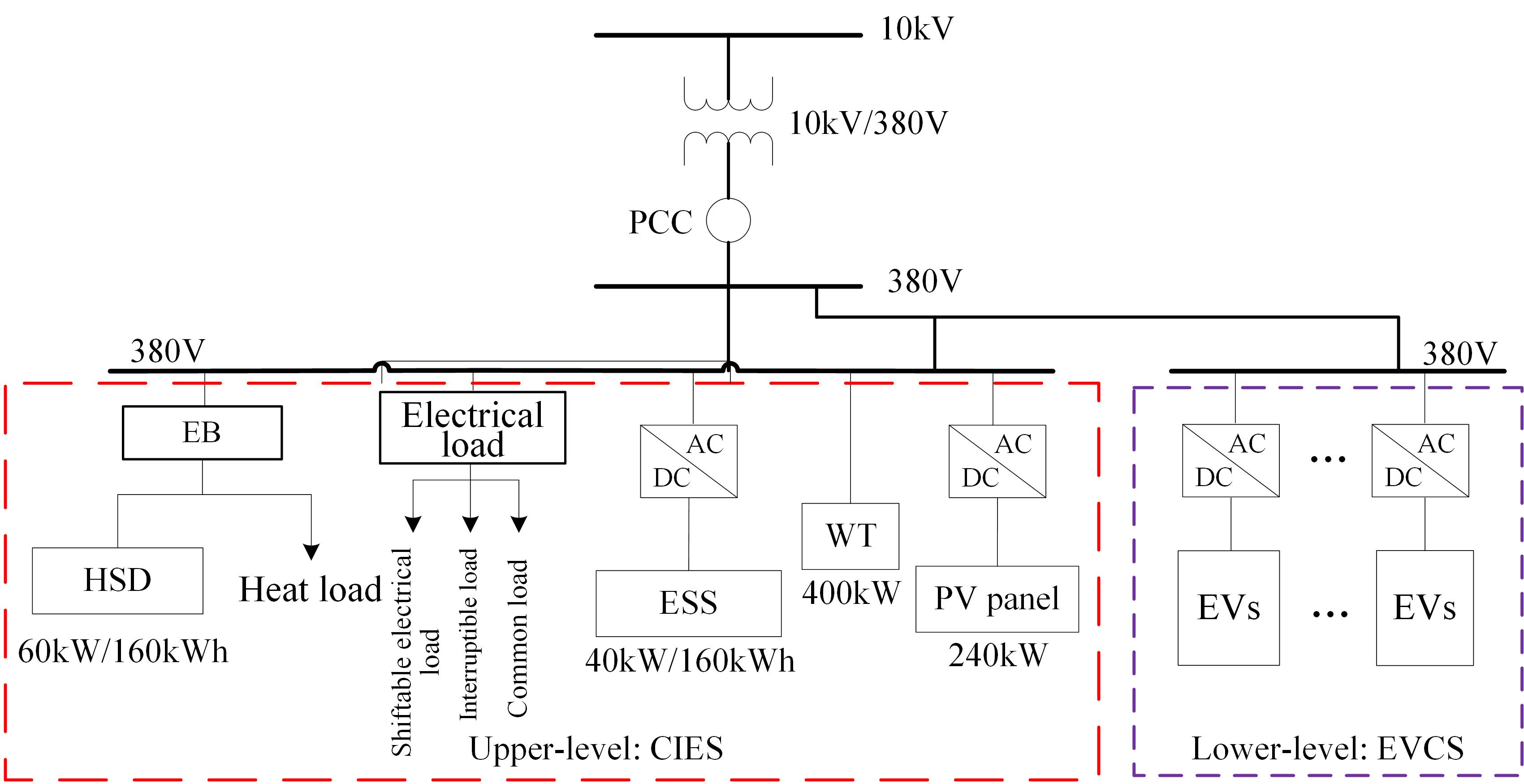}
    \caption{The CIES test system}
    \label{2}
\end{figure}

As illustrated in Fig. \ref{2}, the system includes WT, PV, ESS, HSD, EB, charging piles and EVs. The corresponding parameters are as follows.

(1) Grid parameters: $P_{\max ,t}^{grid}$ =500kW, and the reserve price is 0.04 ¥/kWh; (2) CIES parameters: WT parameters: ${{v}_{in}}=3\text{m}/\text{s}$, ${{v}_{*}}=15\text{m}/\text{s}$, ${{v}_{out}}=25\text{m}/\text{s}$, WT rated power: ${{P}_{r}}=500\text{kW}$; PV module parameters: ${{\eta }_{pv}}=0.093$, ${{A}_{pv}}=3900{{\text{m}}^{2}}$, $P_{\text{max}}^{PV}=360\text{kW}$; ESS parameters: $C_{_{\text{0}}}^{ESS}\text{=}C_{{{\text{T}}_{\text{end}}}}^{ESS}\text{=}C_{_{\text{*}}}^{ESS}\text{=32kW}\cdot \text{h}$, $C_{\text{min}}^{ESS}=32\text{kW}\cdot \text{h}$, $C_{_{\text{max}}}^{ESS}=160\text{kW}\cdot \text{h}$, ${{\eta }_{_{dc}}}\text{=}{{\eta }_{_{ch}}}\text{=0}\text{.9}$, $P_{CH,\max }^{ESS}=P_{DC,\max }^{ESS}=40\text{kW}$; HSD parameters: $C_{\max }^{HSS}=160\text{kW}\cdot \text{h}$, $P_{CH,\max }^{HSS}$=$P_{DC,\max }^{HSS}$=60kW; EB parameters: EB power is 300kW, EB performance coefficient is 0.99; Building parameters: Building comprehensive heat transfer coefficient $K$ is 0.5, Building surface area $F$ is 24000$m^2$; (3) EVCS parameters: The system contains 1 charging station and 10 charging piles, the charging power of each charging pile is 15kW; (4) EV parameters: there are a total of 15 EVs in the CIES, and the battery capacity of each EV is 60kWh, and the charge and discharge efficiency of each EV is 0.9, the total charging power of EVs in a dispatching period is 900kW; (5) The grid TOU electricity prices: refer to Table \ref{tab3} for specific prices at different periods in this community. (6) MCS parameters: sampling times $N$=1000, ${{\mu }_{M}}$ =3.2,  ${{\sigma }_{M}}$ =0.88, ${{\mu }_{1}}$ =17.6, ${{\sigma }_{1}}$=3.4, ${{B}_{c}}$ = 20kWh and ${{E}_{d100}}$ = 15kWh\cite{wei2020aggregation}. 

\begin{figure}[htbp]
\begin{minipage}[t]{0.49\linewidth}
\centering
\includegraphics[height=3.6cm,width=4.5cm]{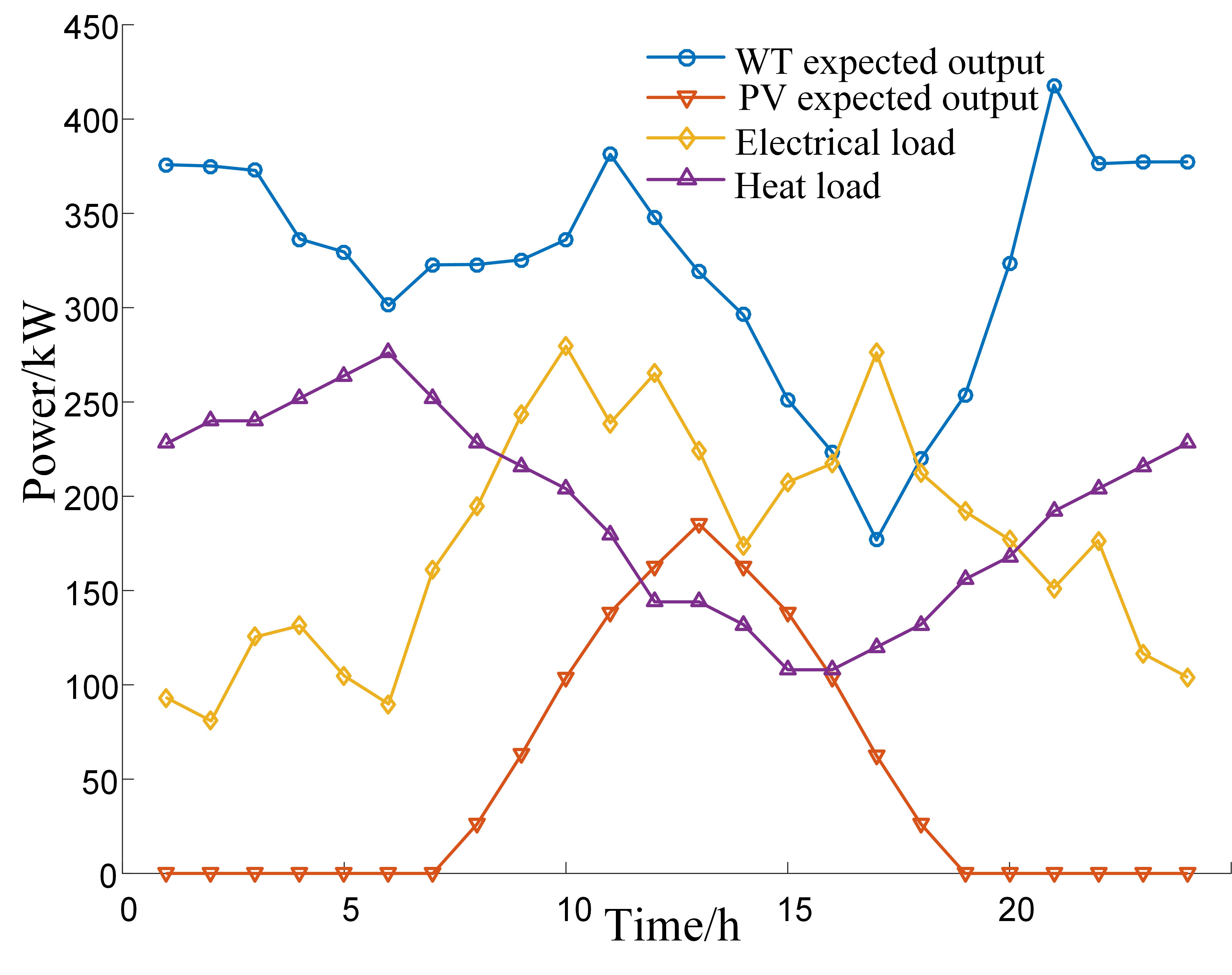}
\caption{WT and PV power output and electricity and heat load demands in different periods}
\label{The WT and PV}
\end{minipage}%
\hfill
\begin{minipage}[t]{0.48\linewidth}
\centering
\includegraphics[height=3.6cm,width=4.5cm]{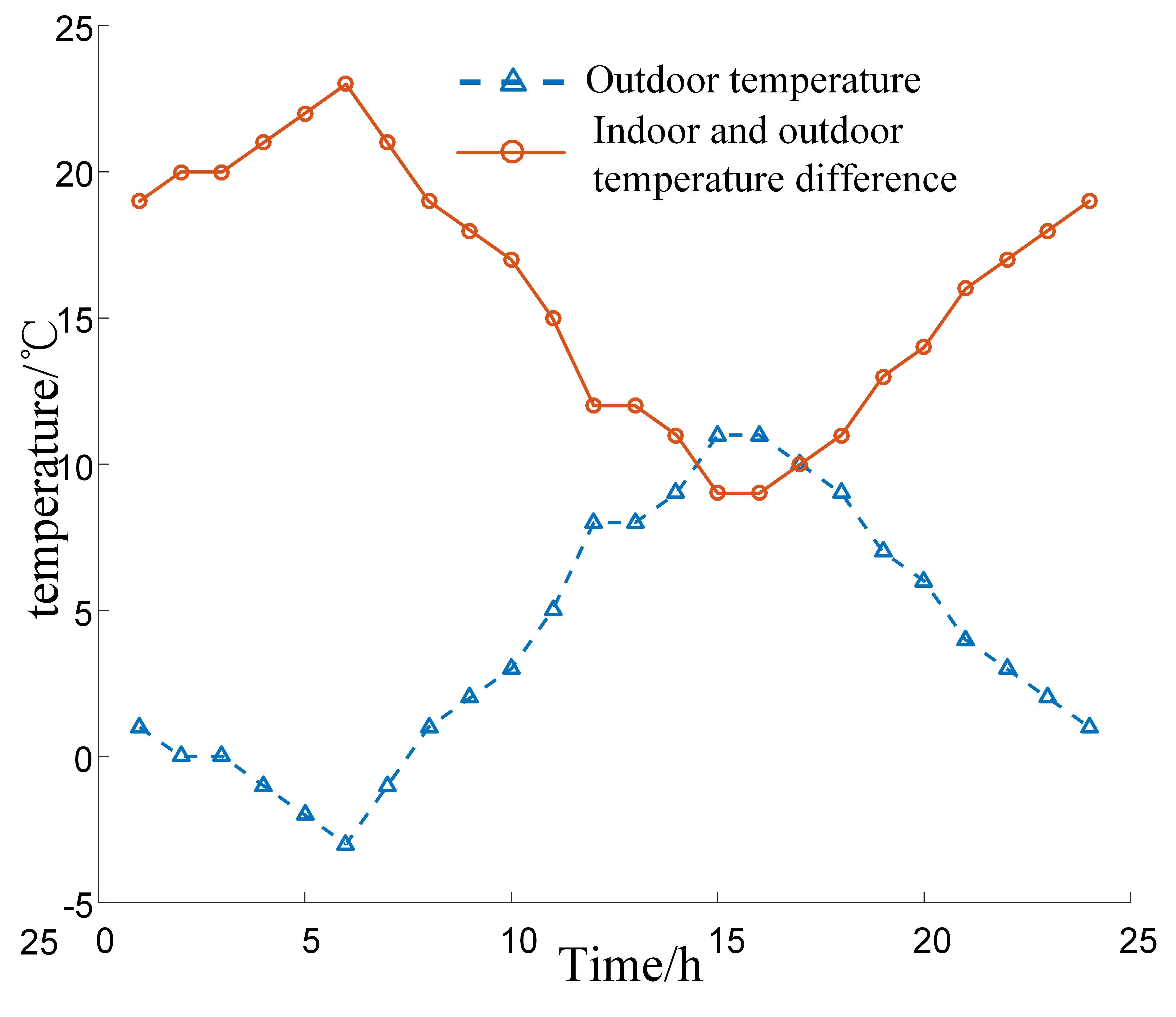}
\caption{Outdoor temperature and temperature difference}
\label{Outdoor temperature}
\end{minipage}
\end{figure}

\subsection{Analysis of Expected RG Outputs and Electricity, Heat and Initial EV Demands }
\begin{figure}[t]
    \centering
    \includegraphics[width=2.2in]{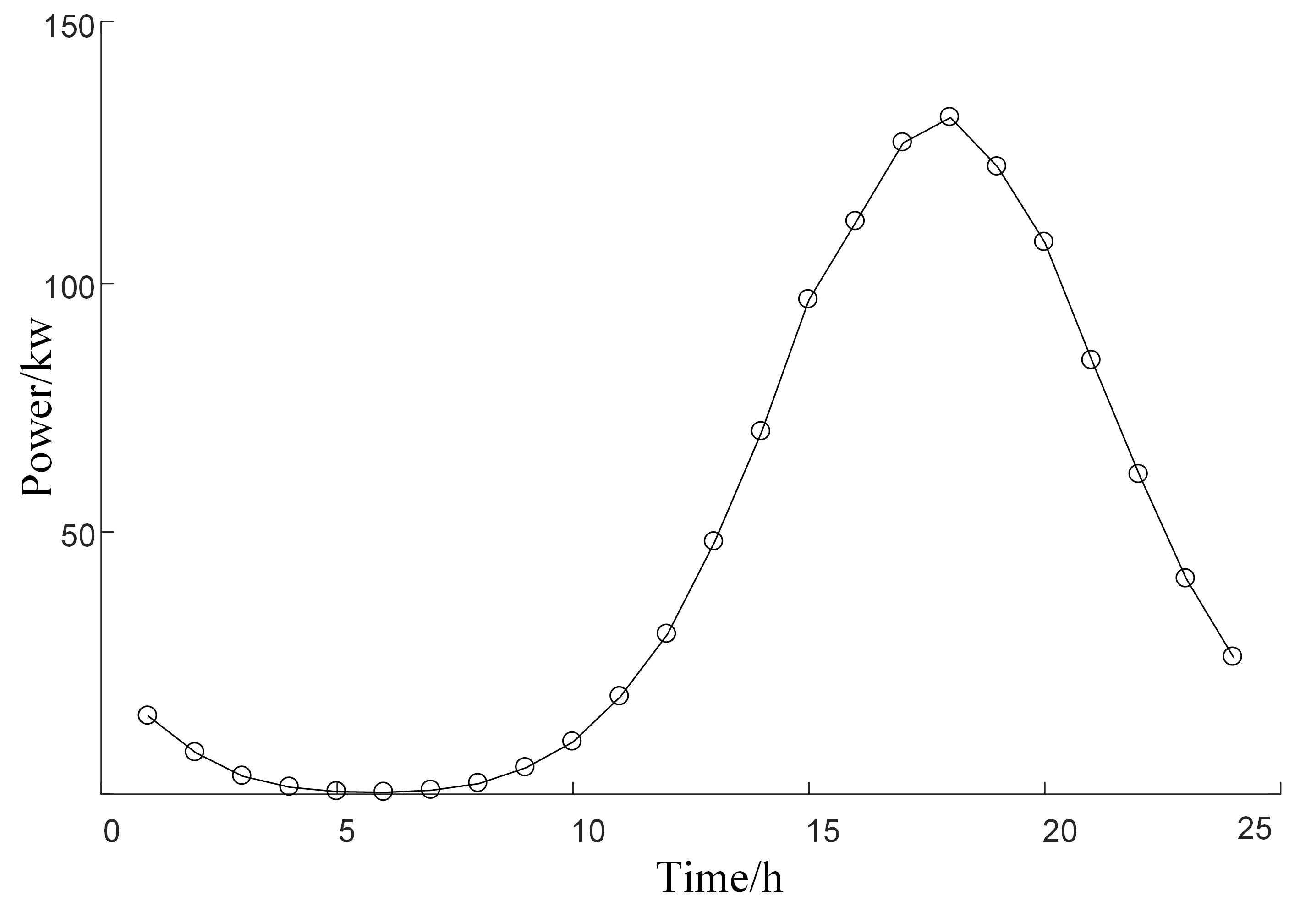}
    \caption{Charging powers of EVs in a disordered state}
    \label{EVdemand}
\end{figure}

Fig. \ref{The WT and PV} shows the WT and PV power outputs, electricity and heat demands of the CIES at different periods. Fig. \ref{Outdoor temperature} shows the outdoor temperature and temperature difference. The indoor temperature is set to a constant temperature 20°C. To meet the indoor temperature demand, the heat load demand in each period is determined by the outdoor temperature. Fig. \ref{EVdemand} illustrates the charging demands of EVs in a disordered state.

\begin{table}[]
\centering
\caption{TOU electricity prices}\label{tab3}
\begin{tabular}{@{}ccc@{}}
\toprule
Periods  & Specific time period  & Electricity prices (¥/kWh)
  \\ \midrule
Peak period  & 8:00-11:00,18:00-21:00  & 0.804  \\
Flat period  & 6:00-7:00,12:00-17:00  & 0.550  \\
  Valley period  & 1:00-5:00,22:00-24:00 & 0.295 \\ \bottomrule
\end{tabular}
\end{table}

\begin{figure}[h]
    \centering
    \includegraphics[width=3.6in]{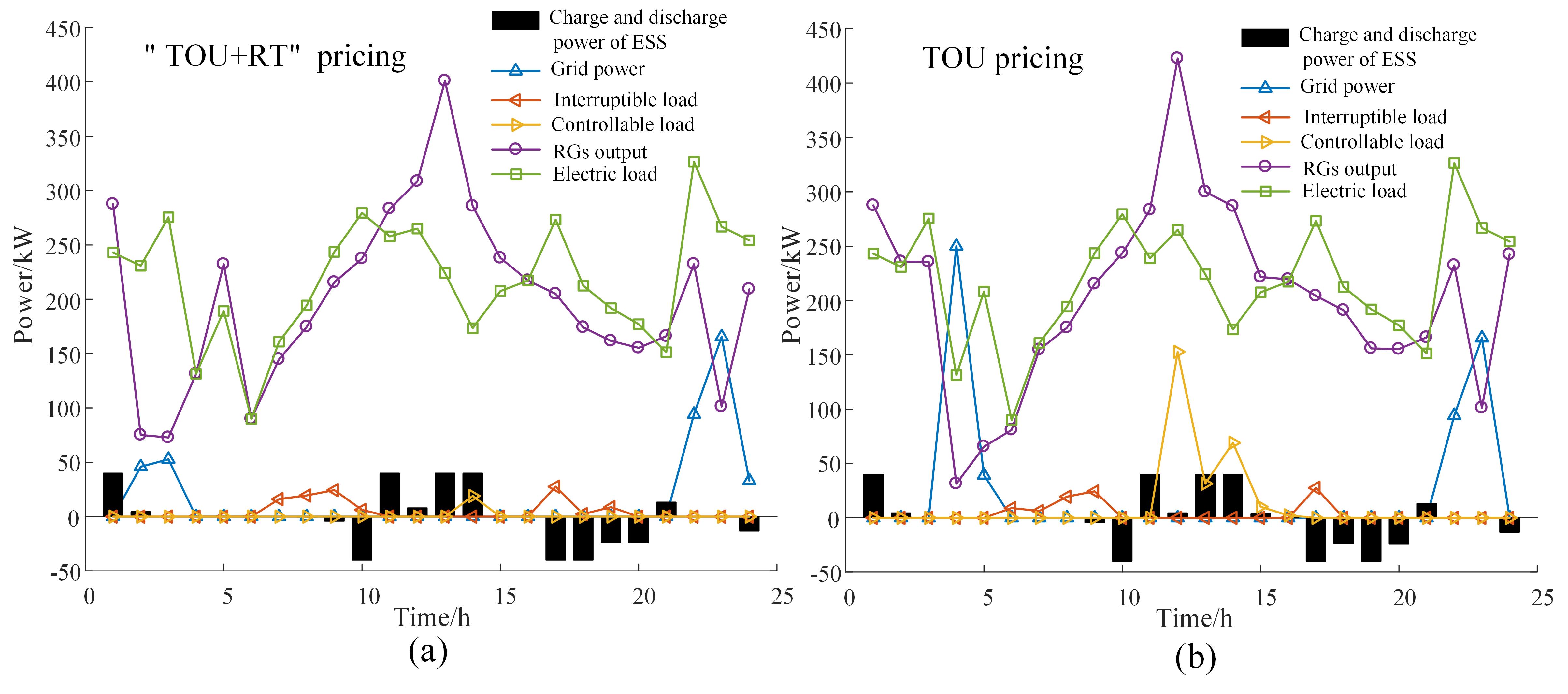}
    \caption{Scheduling schemes of electricity demand under different pricing mechanisms}
    \label{electricity demand}
\end{figure}

\subsection{Comparison of Electricity and Heat Demands Scheduling Schemes under Different Pricing Mechanisms}
\subsubsection{Comparison of Scheduling Schemes for Electricity Demand}

Fig. \ref{electricity demand} shows that  the controllable load is significantly reduced by using the dynamic pricing mechanisms, which indicates that the RGs powers are mainly accommodated by the EVs. By doing so, wind and solar power curtailments can be cut, which indicates that the proposed pricing mechanism is beneficial to accommodate RGs power.

\begin{figure}[!h]
    \centering
    \includegraphics[width=3.6in]{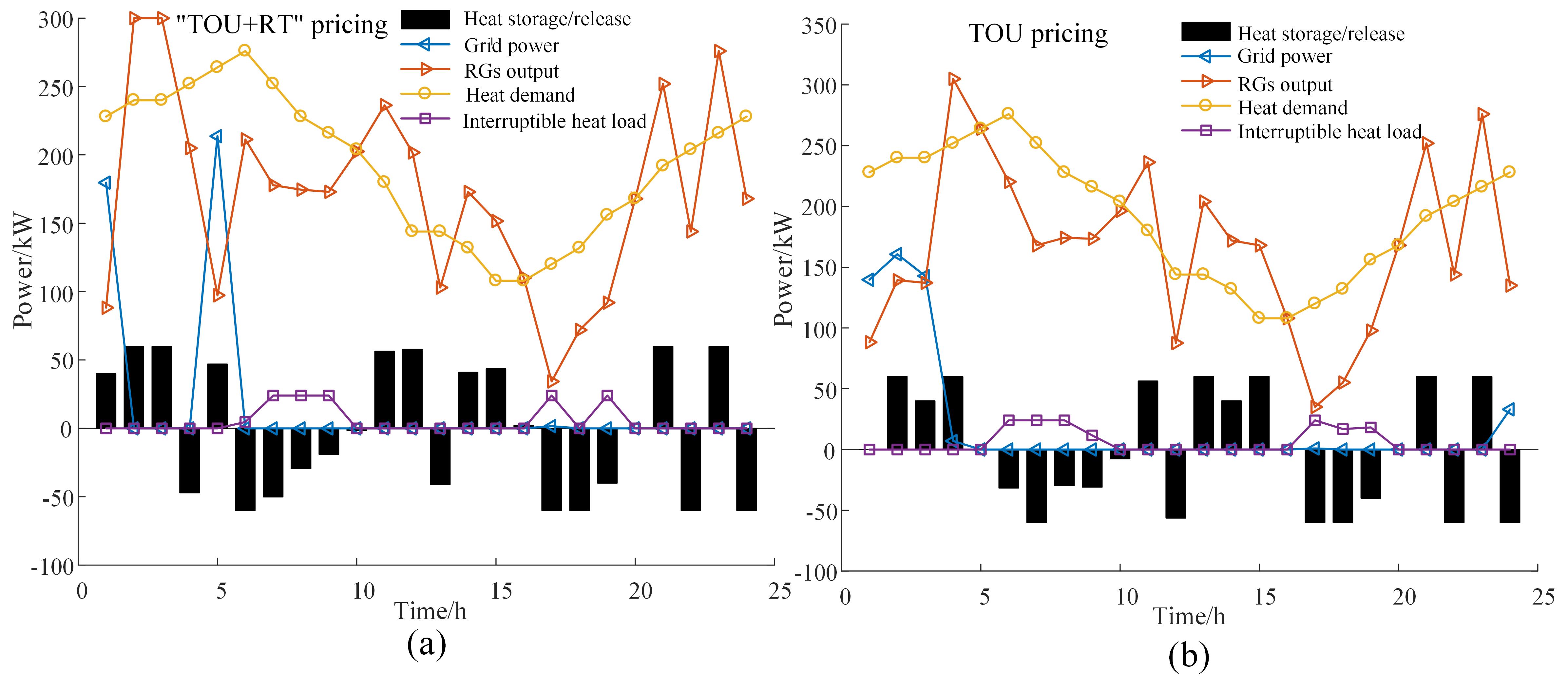}
    \caption{Scheduling schemes of heat demand under different pricing mechanisms}
    \label{heat demand}
\end{figure}
\subsubsection{Comparison of Scheduling Schemes for Heat Demand}
It can be observed from Fig. \ref{heat demand} that in period 1:00-5:00, the “TOU+RT” dynamic pricing mechanism is adopted to decrease the power accommodation of the heat load from the grid. The proposed pricing mechanism can increase the accommodation of renewables while maintaining the balance of supply and demand of heat load. Furthermore, the operating expenses of the CIES are reduced through storing and releasing heat energy of HSD.

\subsection{Comparison of EVCS Scheduling Schemes under Different Pricing Mechanisms}

\begin{figure}[h]
    \centering
    \includegraphics[width=3.6in]{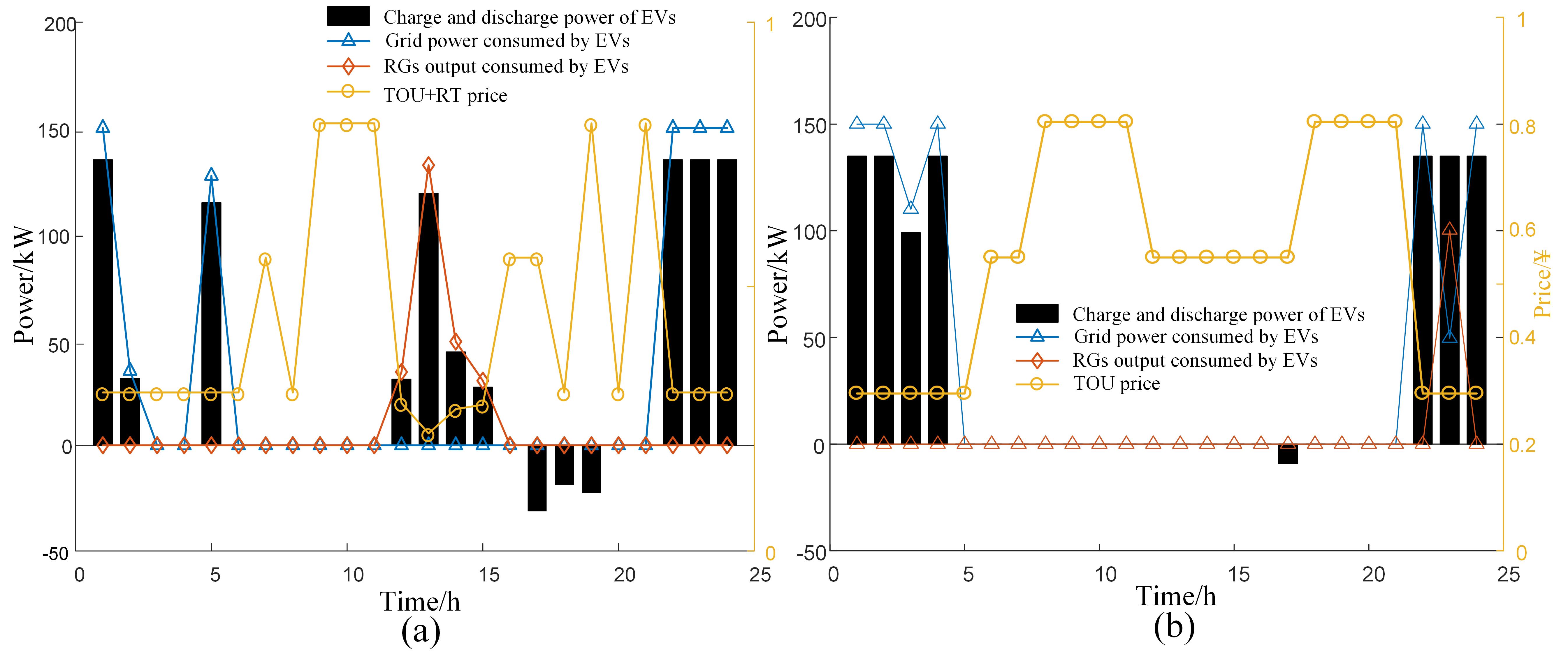}
    \caption{Charging and discharging schemes of the EVCS under different pricing mechanisms}
    \label{EVs}
\end{figure}

Fig. \ref{EVs} illustrates that the electricity price determined by the presented pricing mechanism can enable EVs to more actively participate in the adjustment of the CIES operation than TOU pricing, which can further promote the consumption of the RG outputs. Meanwhile, the proposed pricing mechanism can flexibly lead EVs to properly arrange charging and discharging schemes and avoid centralized charging in off-peak periods. Besides, the period of EVs charging coincides with that of renewable energy accommodation, which can reduce the electricity purchased from grid and increase renewable accommodation.

\subsection{Economic Analysis}
\subsubsection{Economic Analysis of the EVCS Operation Costs}

\begin{figure}[t]
    \centering
    \includegraphics[width=3.6in]{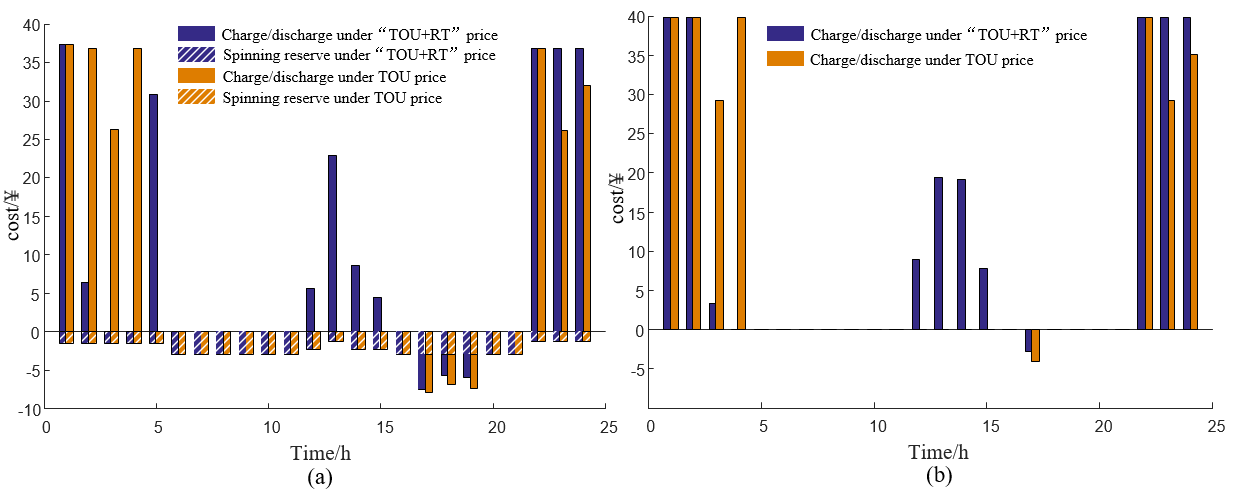}
    \caption{Economic analysis of the EVCS operation costs}
    \label{EVs costs}
\end{figure}  

The EVCS operating costs under different pricing mechanisms and with/without consideration of EVs spinning reserves are illustrated in Fig. \ref{EVs costs}.

In terms of pricing mechanisms, Fig. \ref{EVs costs} (a) and (b) respectively indicate that during most of the periods, the charging-discharging costs of EVs under varied pricing mechanisms are obviously different. This fact suggests that the proposed dynamic pricing mechanism can guide EVs to properly choose charge and discharge periods to achieve flexible demand response. In this way, the total charging-discharging costs of EVCS are reduced.

Regarding EVs spinning reserves, Fig. \ref{EVs costs} shows the impact of EV spinning reserves on the operating cost of the EVCS. Comparing Fig. \ref{EVs costs} (a) and (b), one can see that when EVs participate in the EVCS operation as a spinning reserve provider during most of the periods, the operating cost of the EVCS is significantly lower than its cost when the EVs don't do so. This demonstrates that EVs, as reserve providers, participate more actively in the EVCS operations, which further improves the economy of the EVCS.

\subsubsection{Economic Analysis of the Joint Operation Costs}

Taking the confidence level 90\% as an example, an economic comparison of the operation costs has been made in Table \ref{tabEV}. For one thing, compared with the TOU pricing mechanism, the CIES, EVCS and joint operating costs obtained by using the proposed pricing mechanism are reduced respectively, which verifies the effectiveness and superiority of the proposed pricing mechanism. For another, compared with that not considering EV spinning reserves, the providing of spinning reserves by EVs will bring profits to EVCS and reduces the CIES cost purchasing reserves from the grid. 

\begin{figure}[t]
    \centering
    \includegraphics[width=3.5in]{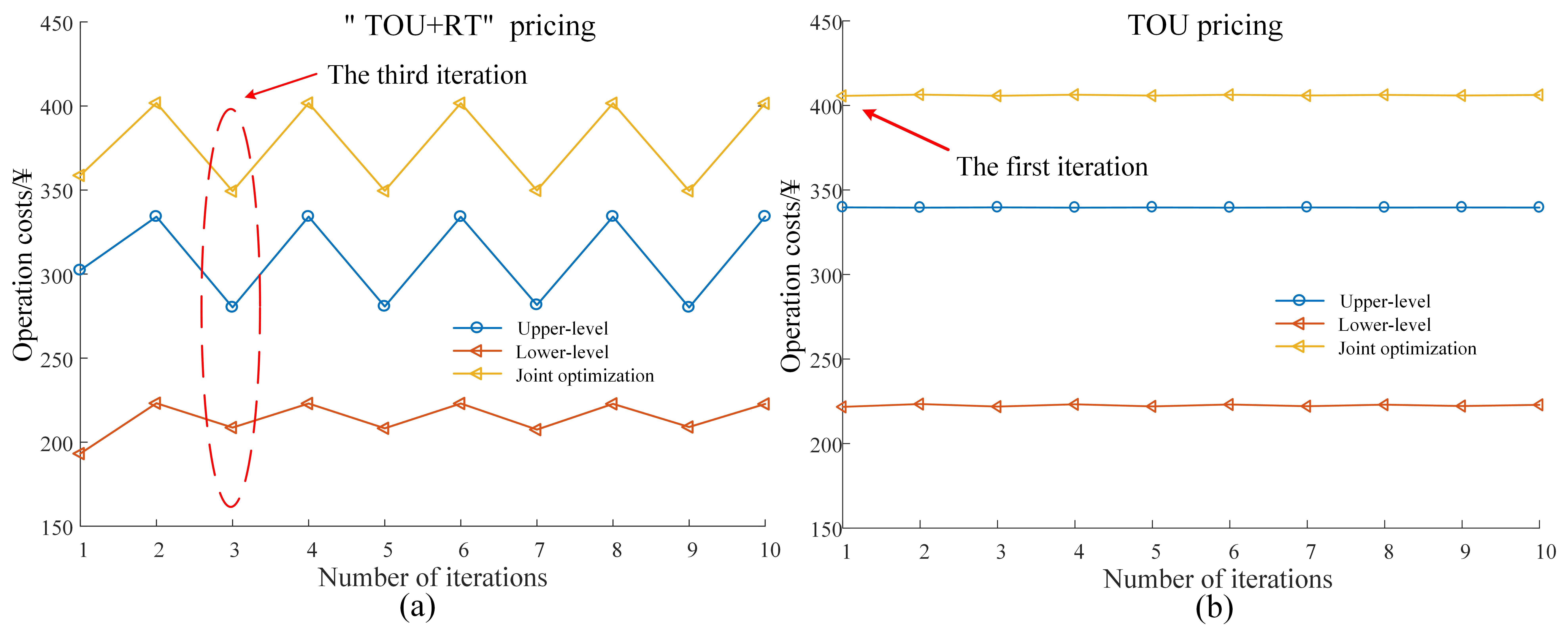}
    \caption{Joint optimal solutions iterative process under different pricing mechanisms}
    \label{Joint optimal}
\end{figure}

\begin{table}[h]
\centering
\caption{Comparison of operating costs with/without considering EVs reserves} \label{tabEV}
\resizebox{250pt}{6.5mm}{

\begin{tabular}{ccccccc}
\hline
Pricing  & \multicolumn{3}{c}{\textbf{Costs considering EVs reserves }}   & \multicolumn{3}{c}{ Costs without considering EVs reserves }        \\  \cline{2-7}   mechanisms
& CIES/¥      &  EVCS/¥         & Joint operation/¥       &  CIES/¥    & EVCS/¥  &  Joint operation/¥  \\ \hline
TOU    & 339.73       & 221.73          & 405.68      &  410.98   & 292.98   & 504.71       \\
\textbf{ “TOU+RT” }       & 280.17            & 208.69    & 349.35          & 349.53   & 281.86   & 449.01    \\ \hline
\end{tabular} }
\end{table}
\subsection{Comparison of Joint Optimal Solution under Different Pricing Mechanisms }

Fig. \ref{Joint optimal} illustrates the iterative solving process of the presented scheduling model under different pricing mechanisms. It shows that the joint optimal solutions of the proposed pricing mechanism and the TOU pricing appear in the third and first iterations, respectively.

\subsection{Flexible Demand Response Analysis}

\begin{figure}[htbp]
\begin{minipage}[t]{0.49\linewidth}
\centering
\includegraphics[height=3.6cm,width=4.5cm]{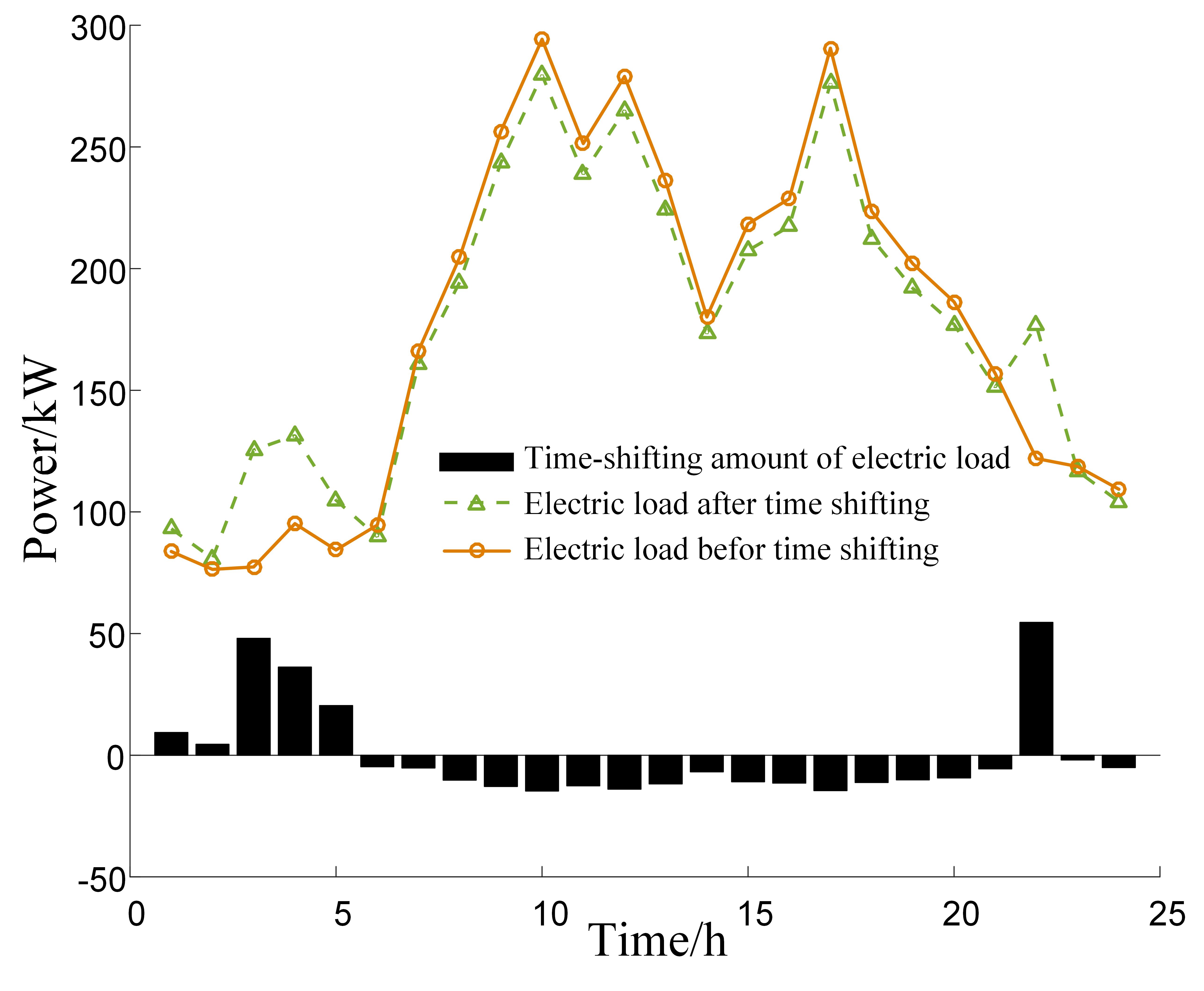}
\caption{Electrical demand before and after time shifting of shiftable load}
\label{Actual power}
\end{minipage}%
\hfill
\begin{minipage}[t]{0.48\linewidth}
\centering
\includegraphics[height=3.6cm,width=4.5cm]{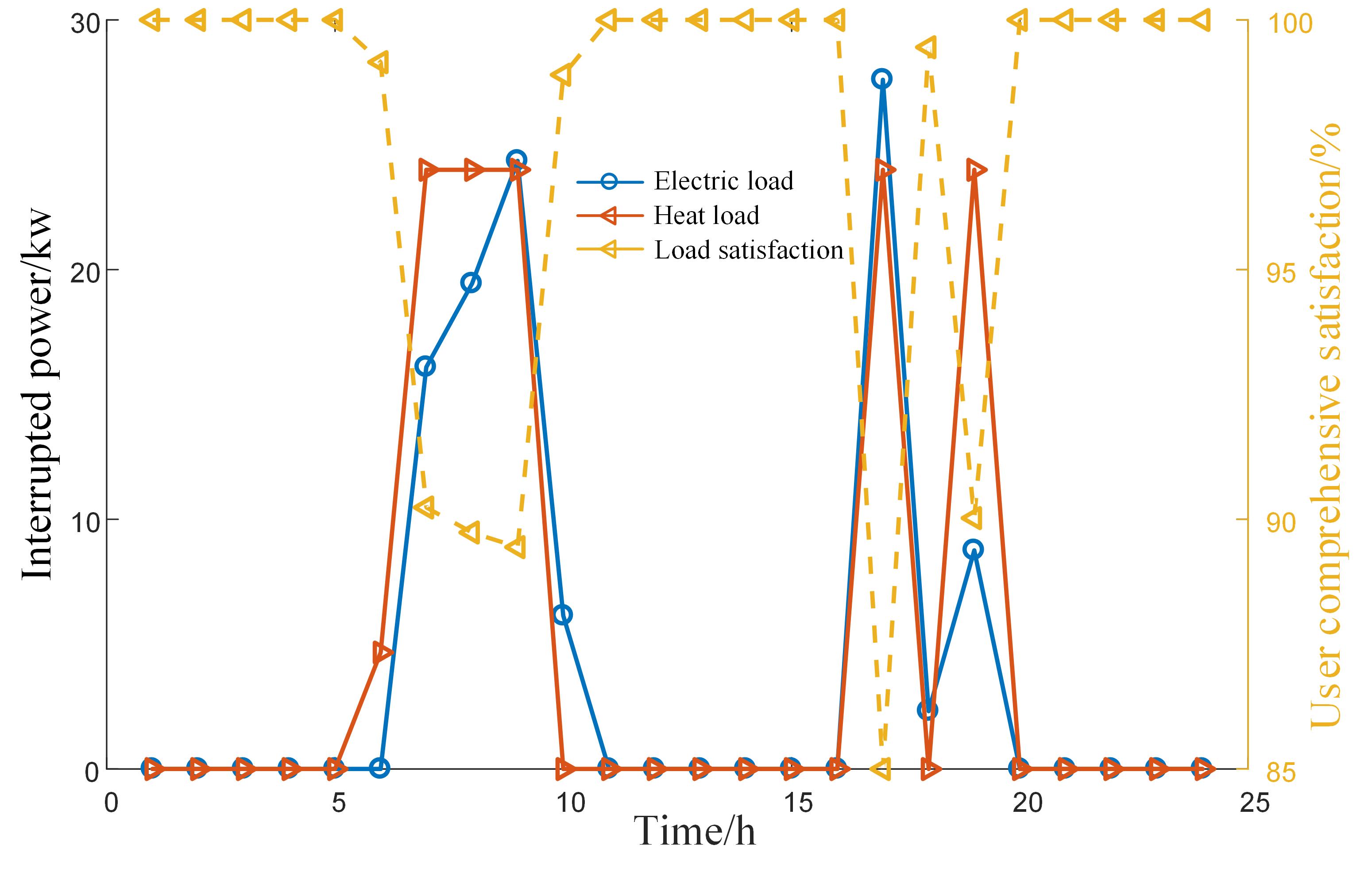}
\caption{Interrupted load powers and user comprehensive satisfaction}
\label{satisfaction}
\end{minipage}
\end{figure}

Fig. \ref{Actual power} illustrates that the electric load is decreased during peak price periods while it is increased during valley price periods. Because of the demand response of shiftable load, the electricity purchased by the system during peak price periods is reduced, which decreases the total CIES operating cost.

Fig. \ref{satisfaction} indicates the change of interruption powers of the interruptible electric and heat load and user consumption satisfaction in a scheduling cycle. One can see that the interruption load works during peak price periods. This phenomenon suggests that users prefer to reduce heating loads by a certain percentage to achieve the goal of reducing cost within an acceptable thermal comfort range. Therefore, interruptible load can achieve the effects of peak shaving, which is beneficial for relieving the power supply pressure of the CIES. Through flexible demand response, the overall operating cost of the system is reduced.

\subsection{ESS and HSD Scheduling Schemes Analysis}

\begin{figure}[htbp]
    \centering
    \includegraphics[width=2.1in]{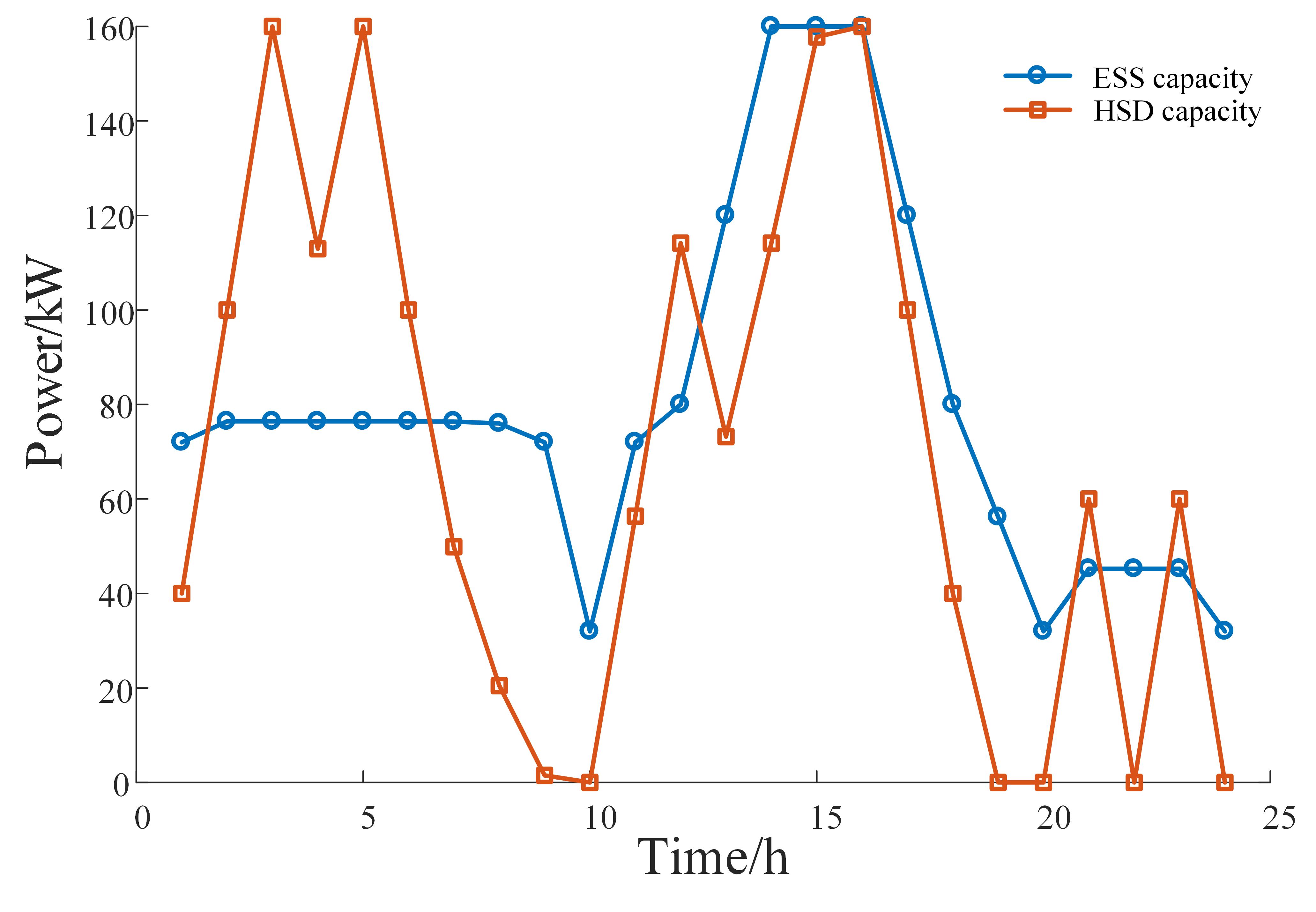}
    \caption{Scheduling schemes of the ESS and HSD }
    \label{ESS and HSD}
\end{figure}

It can be observed from Fig. \ref{ESS and HSD} that during the valley load period, ESS and HSD respectively charge power and store heat, while discharge power and release heat during the peak load period. This verifies that ESS and HSD can play the role of peak shaving and valley filling, effectively alleviating the power supply pressure of the system during peak load periods.

\subsection{Impact Analysis of Different Confidence Levels}

\subsubsection{ Impact of Confidence Levels on the Operating Costs}
\begin{figure}[htbp]
\begin{minipage}[t]{0.49\linewidth}
\centering
\includegraphics[height=3.6cm,width=4.5cm]{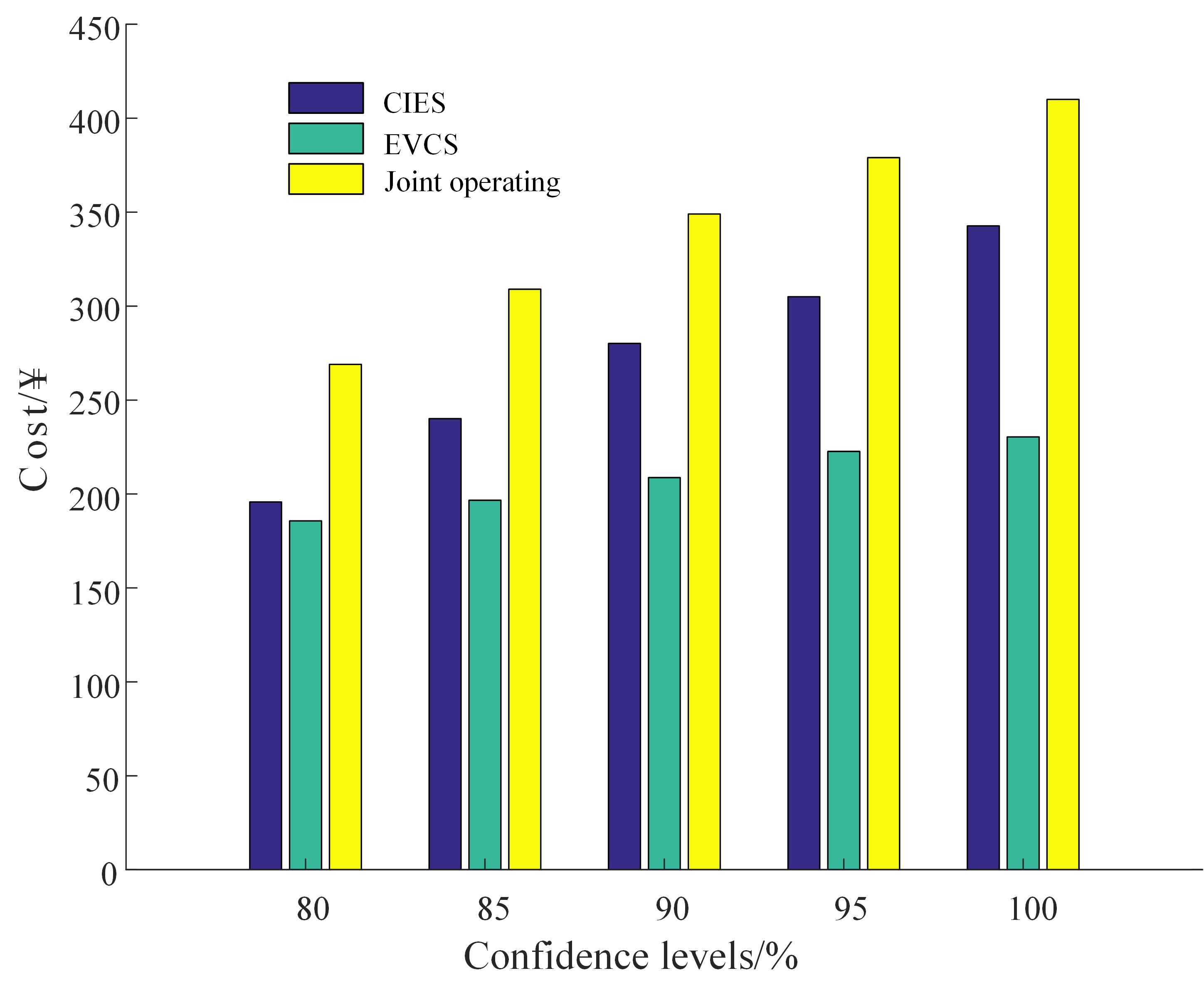}
\caption{Operating costs of CIES and EVCS under different confidence levels}
\label{operating costs}
\end{minipage}%
\hfill
\begin{minipage}[t]{0.48\linewidth}
\centering
\includegraphics[height=3.6cm,width=4.5cm]{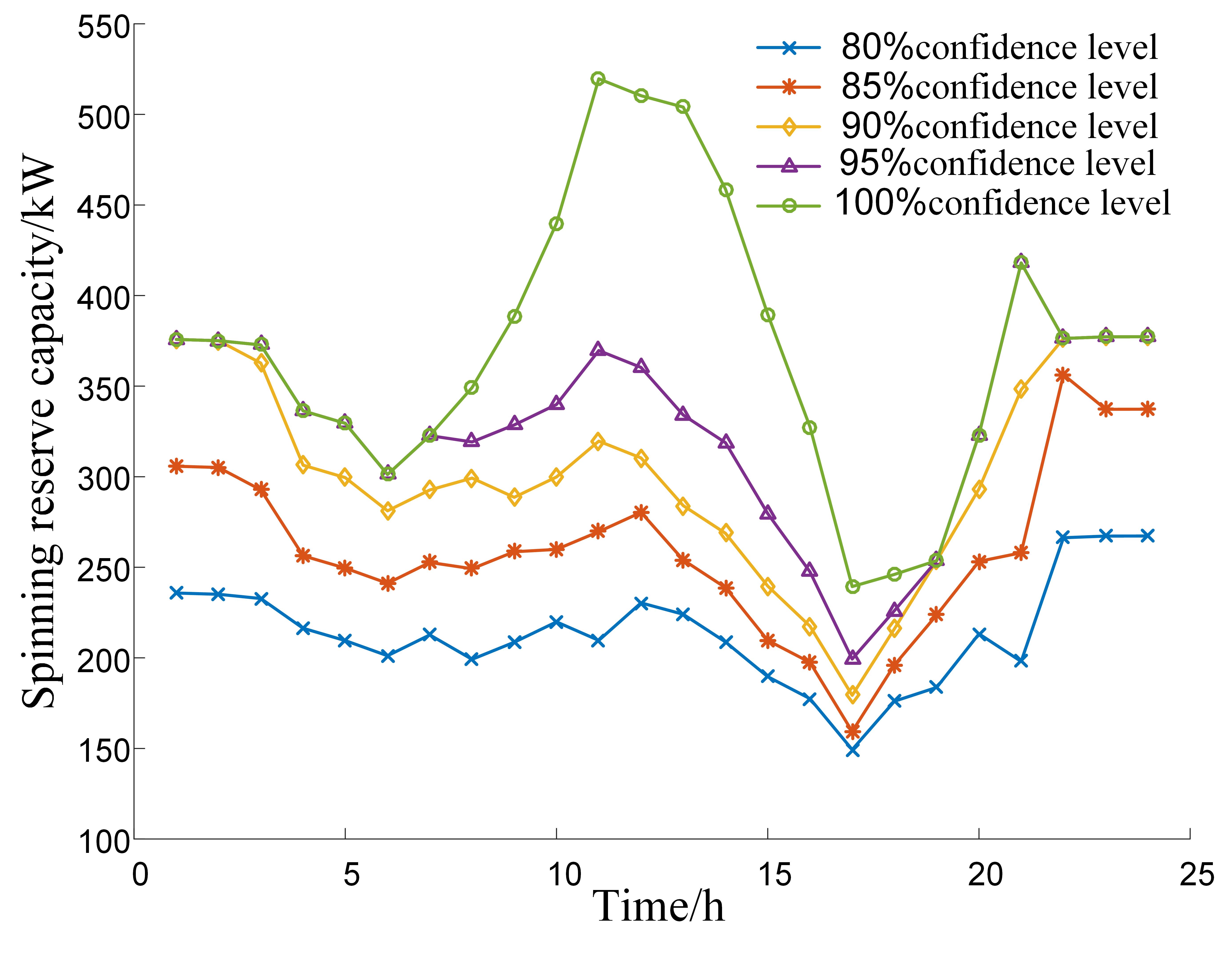}
\caption{Required reserve capacities of the CIES under different confidence levels}
\label{reserve capacity}
\end{minipage}
\end{figure}

Fig. \ref{operating costs} reveals that with the confidence level rising, both the CIES operating cost and the joint operating cost gradually increase. Due to the increasement of the confidence levels, the CIES needs more reserve capacity to keep the balance of supply and demand, resulting in higher operating costs of the CIES.
\subsubsection{Impact of Confidence Levels on Reserve Capacity}

Fig. \ref{reserve capacity} shows that with the increasement of the confidence level, the reserve capacity that the CIES needs accordingly increases, which unavoidably leads to a higher operating cost of the CIES. As a result, it’s of great significance to choose a reasonable confidence level to pursue the balance between economy and reliability of the system operation.

\subsection{Impact of Discrete Steps on the Joint Optimal Costs}

\begin{figure}[!h]
    \centering
    \includegraphics[width=2.5in]{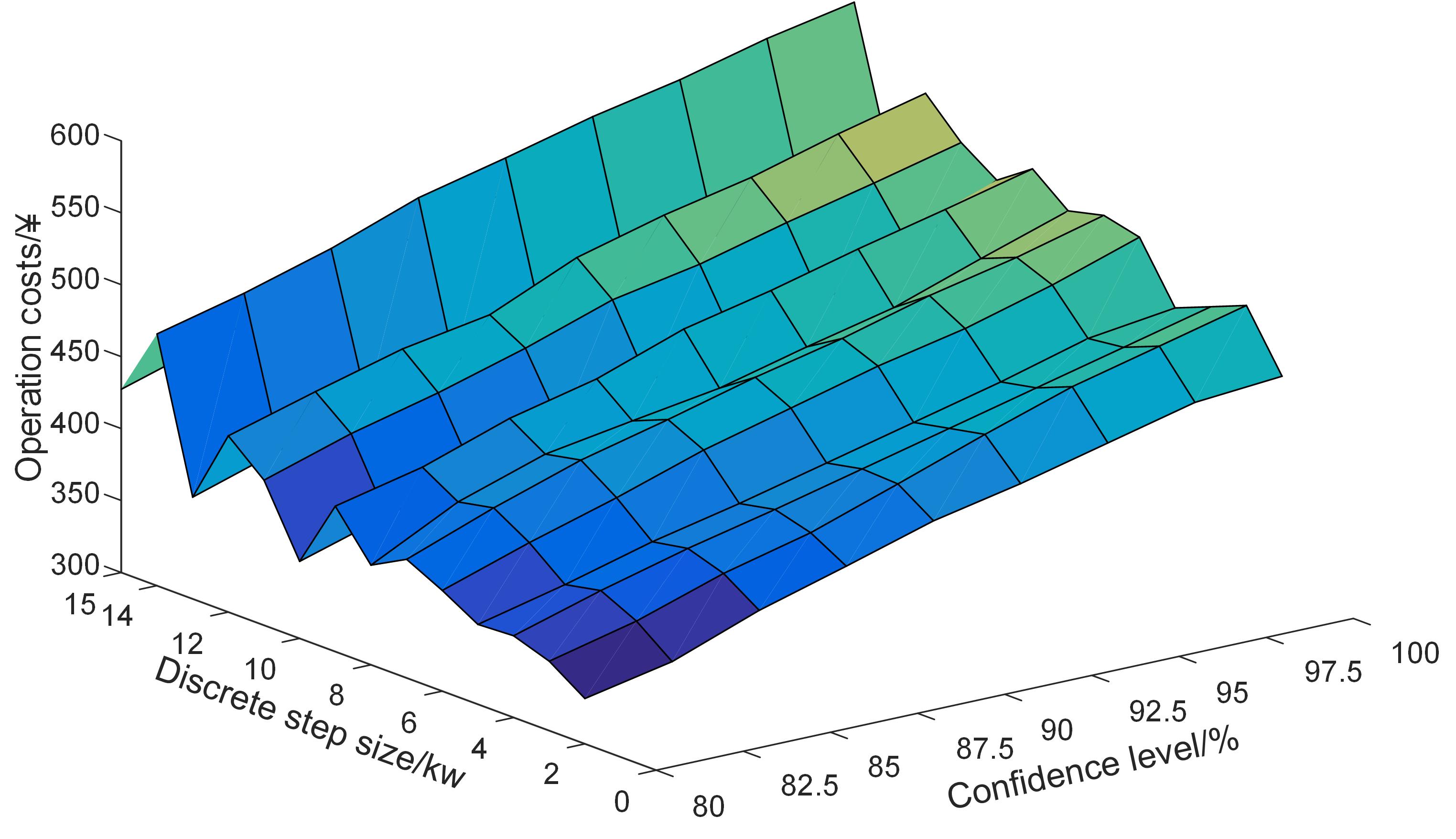}
    \caption{Impact of different discrete steps on the joint optimal costs }
    \label{sanwei}
\end{figure}

Fig. \ref{sanwei} illustrates that when the step size $q$ is greater than 5 kW, the gap between the optimal costs is large at the same confidence level. On the contrary, when the step size $q$ is less than 4 kW, the impact of step sizes on optimization results has been drastically reduced, but the computation time sharply increases. To achieve a compromise between reliability and economy, the appropriate range of the step size is bounded by the interval from 4 kW to 5 kW in this study.

\subsection{Comparative Analysis with Other Algorithms}
To reasonably evaluate the performance of the proposed approach, a comparison with the hybrid intelligent algorithm (HIA) that combines MCS and particle swarm optimization (PSO) has been performed. Here, the parameters of the HIA are set in accordance with \cite{li2018optimal1}.
Considering inherent randomness of the HIA, the average result of 20 independent runs is used as its final result. Given $\alpha $=90\% and the discrete step length $q$=10kW,  the comparison results are shown in Table \ref{tab5}.

\begin{table}[!htbp]
\centering
\caption{Comparison results between the proposed method and HIA} \label{tab5}
\resizebox{250pt}{10mm}{

\begin{tabular}{c|c|c|c|c}
\hline
Confidence & \multicolumn{2}{c|}{The proposed method}        & \multicolumn{2}{c}{HIA}        \\ \cline{2-5}                 level/\% &\begin{tabular}[c]{@{}c@{}}Joint operation\\ costs/¥\end{tabular} & \begin{tabular}[c]{@{}c@{}}Calculation\\ time/s\end{tabular} & \begin{tabular}[c]{@{}c@{}}Joint operation\\ costs/¥\end{tabular} & \begin{tabular}[c]{@{}c@{}}Calculation\\ time/s\end{tabular} \\ \hline
90       &  351.04     & 3.02  & 404.12  & 133.75   \\ \hline
95         & 371.73   & 2.58   & 425.05    & 186.32    \\ \hline
100          & 407.79   & 2.68   & 461.38   & 198.26    \\ \hline
\end{tabular}}
\end{table}

Table \ref{tab5} illustrates that the proposed method outperforms the HIA algorithm from two perspectives: (1) the operating costs are better than that of the HIA under various confidence levels; (2) the calculation time of this method is significantly less than that of the HIA.

\section{Conclusion}\label{Conclusion}
To coordinate flexible demand response and multiple renewable generations uncertainties, this paper proposes a bi-level optimal scheduling model for CIES with an EVCS in multi-stakeholder scenarios, where an integrated demand response program comprising  a  dynamic  pricing  mechanism  is designed.
The simulation results on a practical CIES demonstrate the following conclusions:

(1) The proposed bi-level optimal scheduling model manages  to  balance  the  interests between CIES and EVCS by  coordinating flexible demand response and multiple renewable generations uncertainties.

(2) The designed integrated demand response program can promote a balance between  supply  and  demand while keeping a user comprehensive satisfaction within an acceptable range.

(3) The dynamic pricing  mechanism is able to flexibly guide users’  energy  consumption and EVs’  charge-discharge  behaviors for renewable consumption while reducing  the  joint  operation  costs of CIES and EVCS. Besides, through  EVs’  active  participation  in  providing spinning  reserve  services,  the  economies  of  the CIES  and  EVCS  are  significantly  improved.  

(4) The simulation results on a real-world CIES verify the effectiveness of the presented  method. Furthermore, the proposal outperforms the hybrid intelligent algorithm with better optimization results and higher calculation efficiency.

In future work, power to gas (P2G) will be introduced into the optimal dispatching of CIES in the next step to promote RGs accommodation and multi-energy complementation. Besides, a more realistic solution should  consider a detailed V2G scheduling model.

\bibliographystyle{IEEEtran}
\bibliography{ref}

\end{document}